\documentclass[epsfig, usenatbib]{mn2e}
\usepackage{epsfig}
\usepackage{natbib}
\usepackage{graphicx}
\usepackage{fixltx2e}
\usepackage{multirow}
\usepackage{amsmath}
\usepackage{amssymb}
\usepackage{url}
\title{The stellar evolution of Luminous Red Galaxies, and its dependence on colour, redshift, luminosity and modelling}

\author[Tojeiro et al.]{Rita Tojeiro\thanks{E-mail: rita.tojeiro@port.ac.uk}$^1$,  
  Will J. Percival$^1$, Alan F. Heavens$^2$, Raul Jimenez$^3$\\
$^1$Institute of Cosmology and Gravitation, Dennis Sciama Building, University of Portsmouth,
Burnaby Road, Portsmouth, PO1 3FX \\
$^2$ SUPA, Institute for Astronomy, Royal Observatory Edinburgh, University of Edinburgh, Blackford Hill, EH9 3HJ\\
$^3$ ICREA \& Institute of Sciences of the Cosmos (ICCUB), University of Barcelona, Barcelona 08028, Spain \\
}

\def\gs{\mathrel{\raise1.16pt\hbox{$>$}\kern-7.0pt %
\lower3.06pt\hbox{{$\scriptstyle \sim$}}}}         %
\def\ls{\mathrel{\raise1.16pt\hbox{$<$}\kern-7.0pt %
\lower3.06pt\hbox{{$\scriptstyle \sim$}}}}         %

\begin{document}

\maketitle

\begin{abstract}
We present a series of colour evolution models for Luminous Red Galaxies (LRGs) in the 7th spectroscopic data release of the Sloan Digital Sky Survey (SDSS), computed using the full-spectrum fitting code VESPA on high signal-to-noise stacked spectra. The colour-evolution models are computed as a function of colour, luminosity and redshift, and we do not a-priori assume that LRGs constitute a uniform population of galaxies in terms of stellar evolution. By computing star-formation histories from the fossil record, the measured stellar evolution of the galaxies is decoupled from the survey's selection function, which also evolves with redshift. We present these evolutionary models computed using three different sets of Stellar Population Synthesis (SPS) codes. We show that the traditional fiducial model of purely passive stellar evolution of LRGs is broadly correct, but it is not sufficient to explain the full spectral signature. We also find that higher-order corrections to this model are dependent on the SPS used, particularly when calculating the amount of recent star formation. The amount of young stars can be non-negligible in some cases, and has important implications for the interpretation of the number density of LRGs within the selection box as a function of redshift. Dust extinction, however, is more robust to the SPS modelling: extinction increases with decreasing luminosity, increasing redshift, and increasing $r-i$ colour. We are making the colour evolution tracks publicly available at \url{http://www.icg.port.ac.uk/~tojeiror/lrg_evolution/}.
 \end{abstract}

\begin{keywords}
galaxies: evolution - cosmology: observations - surveys
\end{keywords}

\title{Modelling LRG evolution}

\section{Introduction}  \label{sec:intro}


In the currently favoured model of galaxy formation, structure builds up hierarchically and galaxies build up their stellar mass within host dark matter haloes via a complex mixing of two processes - star formation from cold gas, and merging. This model is supported by observed large-scale structure and its evolution, which can be well understood and remarkably well modelled within the $\Lambda$ Cold Dark-Matter scenario (LCDM). Because of the complicated astrophysics involved, the link between mass build-up and star-formation is not direct. There are two main observables with which to understand this link: stellar light, which holds information about the formation history of the stars present in any given galaxy at the time of observation, and direct observation of size, density and clustering of individual galaxies and their evolution with redshift. The combination of these two observables has the potential to resolve a long-standing quest of galaxy evolution - to match samples of galaxies at different redshifts, without making assumptions about their stellar or dynamical evolution.

Luminous Red Galaxies (LRGs), for which typically the label of Early-Type Galaxy (ETG) also applies, occupy an interesting position within the galaxy evolution puzzle. They occupy and dominate the high-mass end of the galaxy mass function and are predicted to occupy the most massive dark matter halos. According to LCDM, these halos must have started their assembly earlier, and such a signature is indeed seen in the galaxy stellar populations - LRGs are clearly dominated by old stars, with 70-90 per cent of the present stellar mass having been formed at $z > 1$, with the detailed measurement dependent on the study and exact sample \citep{WakeEtAl06, BrownEtAl07, CoolEtAl08}. \cite{vanDokkumEtAl10} have extended this to higher redshift, and measured growth of massive galaxies - albeit not exclusively red galaxies - as factor of 2 since $z=2$. They concluded this growth was mostly due to mergers, rather than star formation. There is also evidence that more massive LRGs have, on average, older stellar populations, giving strength to the hierarchical formation scenario (e.g. \citealt{ThomasEtAl05, JimenezEtAl07}). What is not clear, however, is why star-formation seems to have shut down in these massive galaxies since roughly a redshift of two (see \citealt{PengEtAl10} for a recent proposal, and references within). The low-level of recent star formation suggests that these galaxies have either been dynamically passively evolving, or that any mass build up through merging in the recent Universe must have occurred through either dry merging (resulting in no new star-formation), or minor mergers (perhaps inducing only very small levels of star formation). It is also possible that the recent star formation observed in LRGs is due completely or in part to the gas returned by stars in the giant branch (AGB) to the interstellar medium. Given the low levels of star formation activity in these galaxies, such a source of gas would provide a natural explanation for the low level of star formation activity observed in LRGs at $z<1$.

Measuring the mass build-up of massive galaxies is therefore a crucial and stringent test of models of galaxy formation. In spite of being a puzzle in itself, the process that shut down star formation in massive early-type galaxies has made this particular population extremely attractive for observational cosmologists. The fact that they are luminous and are thought to have a simple stellar evolution make them easy to target and observe at a range of redshifts. The ability to predict the colour evolution as a function of redshift ensures an efficient allocation of fibres or slits for spectroscopic measurements, and the fact they are very massive makes them observable to larger distances. Finally, LRGs are often approximated as being a uniform population of galaxies that experiences no significant amount of merging. This is attractive as the computation of their predicted density and velocity bias evolution then becomes trivial \citep{Fry96}, enabling more information to be recovered using the clustering of this population to form a standard ruler to measure cosmological geometry or using them to measure Redshift-Space Distortions.

There is a wealth of literature on LRGs and ETGs, their properties and evolution. Studies have been performed based on (see also references within): the mass or luminosity function \citep{WakeEtAl06, BrownEtAl07, FaberEtAl07, CoolEtAl08}; colour-magnitude diagram \citep{CoolEtAl06, BernardiEtAl10b}; photometry SED fitting \citep{KavirajEtAl09, MarastonEtAl09}; absorption line fitting to individual galaxies' spectra \citep{TragerEtAl00, ThomasEtAl05, ThomasEtAl10, CarsonEtAl10} or to stacked spectra \citep{EisensteinEtAl03, GravesEtAl09, ZhuEtAl10}; full spectral fitting \citep{JimenezEtAl07}; close-pair counts \citep{BundyEtAl09} and clustering \citep{ShethEtAl06, MasjediEtAl06, ConroyEtAl07, WhiteEtAl07, MasjediEtAl08, WakeEtAl08, TojeiroEtAl10, deProprisEtAl10}. 

These papers can be broadly split into two types: those that aim to gain knowledge of the stellar content and evolution of these galaxies, and those that are interested in their dynamical evolution, or merging history. The picture seems set in its broad terms, but there is disagreement in the details. It is well established, from the work cited above and many more, that ETGs constitute a very uniform population of galaxies; are dominated by old and metal rich stellar populations; their mean ages (either mass- or light-weighted) decrease with luminosity; and the most luminous occupy more dense environments. There is, however, an increasing amount of evidence pointing towards some amount of recent star formation in intermediate-mass ETGs coming, e.g. from UV excess measurements \citep{KavirajEtAl07,SalimEtAl10}. This amount of star formation is not in conflict with the hierarchical model of structure formation, and \cite{KavirajEtAl10}, through evidence coming from small morphological disruptions in early-type galaxies, argue that it can be explained from the contributions from minor-mergers. The overall dynamical evolution is best constrained via a clustering analysis, although this traditionally involves assuming a model for the stellar evolution of the galaxies in the sample. Measurements vary (see Table 4 in \citealt{TojeiroEtAl10} for a summary), but luminosity growth seems to be confined to less that 20 per cent since a redshift of 1.

\subsection{This work}

The motivation for this work is two fold. In \citet{TojeiroEtAl10} we constrained the dynamical evolution of Sloan Digital Sky Survey (SDSS) LRGs by assuming the stellar evolution of these galaxies was known, and followed the passive evolution model (with a small metal-poor component) of \citet{MarastonEtAl09} (M09). Given the evidence that ETGs have different histories according to their luminosity, as well as evidence for recent star formation in at least intermediate-mass ETGs, the assumption of passive evolution after a single star burst for {\em all} LRGs is clearly an oversimplification. The first of our goals is therefore to provide the community with empirically derived colour evolution models for the SDSS LRG sample that depend on luminosity, colour and redshift.

Secondly, our approach of using the fossil record to infer their star-formation and metallicity histories, as well as dust content, and from there compute their colour evolution as a function of redshift makes minimal prior assumptions about the evolution of the galaxies. The method adopted is to compute the star-formation and metallicity histories using VESPA \citep{TojeiroEtAl07} - a full-spectrum fitting code that assumes no prior information about the star-formation or metallicity history of a galaxy other than it is non-negative. Non-parametric full-spectral fitting has advantages and disadvantages when compared to other methods. The main advantage comes from a well time-resolved and completely unconstrained star formation history, which does not suffer from any problems and biases traditionally associated with SSP-equivalent ages and analyses (e.g. \citealt{TragerEtAl09}). On the other hand, full-spectral fitting can be more sensitive to inaccuracies in the modelling (particularly dust), and spectrophotometric calibrations. Parameter degeneracies are not a problem, in the sense that they can be estimated and incorporated into any analysis.

There are also clear advantages in using the fossil record to compute the colour evolution of a sample of galaxies. Primarily, one does away with the circular methodology of pre-selecting a sample across a range of redshifts that one believes is itself a uniform population of galaxies.  This is traditionally done according to an evolutionary model and after applying the necessary K+e corrections and survey selection functions, and culling the observed sample to one that is self-consistent (i.e., any galaxy in this culled sample could have been observed at any redshift probed by the survey, according to a model). The obvious danger of such an approach is that the deductions about the evolution of the sample simply reflect the assumptions used to define it. In our approach, however, evolutionary colour paths that stray from the colour selection box {\em are} acceptable. We therefore de-couple the sample of galaxies observed at any given redshift from galaxies observed at earlier epochs, and we are insensitive to changing survey selections. Of course, if one wants to interpret this evolution in terms of the overall evolution of a population of galaxies then sample selection becomes crucially important. We leave that for a series of follow-up papers. In this work we introduce a new set of empirical models for the evolution of galaxies labelled as Luminous Red Galaxies, as a function of their observed colour, luminosity and redshift.

This paper is organised as follows. We start by describing the dataset used in this work in Section~\ref{sec:data}; in Section~\ref{sec:method} we describe in detail the several steps of our methodology; in Section~\ref{sec:models} we describe the three sets of SPS models used throughout the paper; we follow by presenting our results in Section~\ref{sec:results}, which we interpret in Section~\ref{sec:model_dependence};  and we finally summarise and conclude in Section~\ref{sec:conclusions}.

\section{Data} \label{sec:data}

The SDSS is a photometric and spectroscopic survey, carried from a dedicated 2.5m telescope in Apache Point, New Mexico. Photometry was taken in five bands: $u, g, r, i$ and $z$, corresponding to central effective wavelengths of 3590\AA , 4819\AA , 6230\AA , 7640\AA , and 9060\AA\ respectively. For details on the hardware, software and data-reduction see \citet{YorkEtAl00} and \citet{StoughtonEtAl02}. In summary, the survey is carried out on a mosaic CCD camera \citep{GunnEtAl98}, two 3-arcsec fibre-fed spectrographs, and an auxiliary 0.5m telescope for photometric calibration. 

Objects were selected for spectroscopic follow-up according to two main targeting algorithms. The main galaxy sample \citep{StraussEtAl02} is a magnitude-limited, high-completeness ($>$99\%) sample, selected in the $r-$band. The targeting is done down to $r=17.77$, resulting in a median redshift of around $\bar{z}=0.11$. The luminous red galaxies sample extends the redshift range to $0.15 <z < 0.5$ by targeting luminous and very red objects according to the target algorithm described in \citet{EisensteinEtAl01}.

\begin{figure}
\includegraphics[width=3.5in]{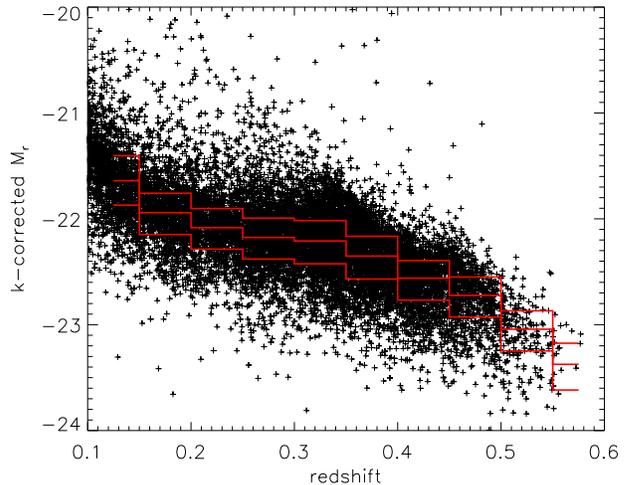}
\caption{The redshift - absolute magnitude diagram of the DR7 LGRs used in this paper (random subsample of 20,000 objects). The magnitudes are rest-frame $r-$band absolute magnitudes, and are k-corrected only. The red lines show the limits that split our sample into 4 magnitudes bins at each redshift (see Section~\ref{sec:data} for more details). k-corrections were calculated using the k-correct code of \citet{BlantonEtAl07}.}
\label{fig:Mr_limits}
\end{figure}

\begin{figure}
\includegraphics[width=3.5in]{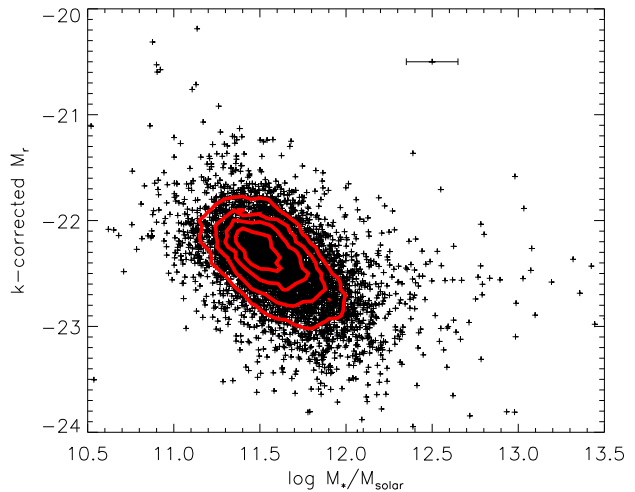}
\caption{Absolute magnitude vs. stellar mass for the LRG sample. The stellar masses are from the public VESPA database (\citealt{TojeiroEtAl09}). The error bar at the top-right corner shows a typical error on the stellar mass. }
\label{fig:mass_Mr}
\end{figure}

In this paper we analyse the latest SDSS LRG sample (data release 7, \citealt{AbazajianEtAl09}), with a spectroscopic footprint of around 8000 sq. degrees and which includes around 180,000 objects. Fig.~\ref{fig:Mr_limits} shows the $r-$band absolute magnitudes as a function of redshift for a random subsample of our galaxies. The k-corrections were calculated using the k-correct code of \citet{BlantonEtAl07}. Strictly, to be self-consistent and completely model independent, we should derive k-corrections from the observed spectra. However, the difference between this approach and using the model fits of the k-correct package will be small, and this will not significantly affect our analysis.

The goal of our work is to provide independent evolutionary histories for samples of galaxies. Luminosity evolution corrections are therefore a result from our work, and cannot be assumed a-priori. We therefore choose samples based only on k-corrected luminosities, with limits designed to optimise our sampling at each redshift, such that the number of galaxies in each luminosity range is kept approximately constant. We show these boundaries in Fig.\ref{fig:Mr_limits}.

In Fig.~\ref{fig:mass_Mr} we show the stellar masses of the LRGs in our sample, and how they relate to the absolute magnitudes. The scatter in this relationship is dominated by uncertainties in the stellar mass estimation, but they are not sufficient to explain all of the observed scatter. The remaining stochasticity indicates intrinsic differences in star formation histories and mass-to-light ratios between different LRGs.

\section{Method} \label{sec:method}

Our methodology can be summarised in the following series of steps:
\begin{enumerate}

\item we stack the observed spectra in regions of colour-luminosity-redshift space, in order to obtain a consistent signal-to-noise ratio (SNR) across this plane;
\item we analyse the stacked spectra with VESPA, in order to obtain a star-formation history (SFH), a metallicity history and dust content representative of the galaxies in any given cell of colour-luminosity-redshift space;
\item we estimate systematic errors due the limitation of the models, so that we can later study their impact on the recovered parameters;
\item we smooth the recovered SFH in lookback time, to obtain a continuous function with time; and
\item we evolve the observed spectrum with lookback time, to obtain a rest-frame spectrum as a function of time from which we can compute any suite of observed-frame colours.

\end{enumerate}

In the rest of this section we explain in detail each of these steps.

\subsection{Stacking the data}

We stack galaxies of similar colour, luminosity and at the same redshift. The resulting spectrum gives a weighted evolution of all the galaxies within the stack.

We work on a fixed grid in redshift, from 0.15 to 0.5 in steps of 0.05. At each redshift, we split the sample into four luminosity slices, such that each slice has approximately the same number of galaxies. Within each redshift bin and luminosity slice, we split the galaxies into stacks based on $r-i$ colour, with the number of stacked objects chosen such that each stack contains a fixed number of objects, set to be around 1000\footnote{For the sake of clarity, we will continue to refer to each region bound by colour, luminosity redshift as a cell, to each column of cells for the same redshift as a redshift bin, and to each grid of colour-redshift for a given luminosity range as a luminosity slice.}. This number was chosen so that the SNR of the stacked spectra would permit differentiation between ages of interest in this paper (see Section \ref{sec:age_grid} for details). The typical photometric errors increase steeply into the blue, so the spread in $g-r$ colour of LRGs would be expected to be larger than that in $r-i$ even if LRGs did not have any intrinsic scatter in their colours. There is also a linear relationship between redshift and $g-r$ up to $z \lesssim 0.35$, which we capture by binning in redshift. We therefore stack in $r - i$ colour rather than $g-r$.
The result is a number of non-uniform grids (one per luminosity slice) in $r-i$ colour, depending on the density of objects in the colour-redshift plane. We choose to do this in favour of a uniform grid in order to keep a roughly constant SNR of the resulting stacked spectra. The disadvantage is that sparse regions in these diagrams need to be larger in order to reach the target number of galaxies, and variations in the intrinsic nature of the galaxies within each cell may be larger. This does not invalidate the resulting evolution model for this cell, which will simply be a roughly luminosity-weighted average evolution for all the galaxies within the cell, although it may mask larger departures from the mean within the cell. 
The above procedure occasionally results in cells that are smaller, in $r-i$, than the typical error on $r-i$ for {\em individual} galaxies. This means that there is likely scatter of galaxies across cells, and that the VESPA solutions from adjacent cells should only change smoothly (or as smoothly as the observational uncertainty in whatever quantity we bin). In other words the true resolution in $r-i$ is given by the photometric errors, and not by the size of the cells.

The goal of this paper is to compute a set of empirical evolutionary models for the {\em colour} of LRGs, and once again we assume that within a cell this is independent of luminosity. This assumption is strengthened by us taking more than one luminosity range, but there may be residual dependencies within a given chosen range. The purpose of the stacking procedure, therefore, is to compute an optimal estimate of the underlying spectrum, of which we assume each galaxy represents a random realisation. We discard luminosity information within each cell by normalising all spectra to a common wavelength, and we compute an inverse-variance weighted average, per wavelength point:

\begin{equation} \label{eq:stack_flux}
F \equiv \langle f_\lambda \rangle = \frac{\sum_i f_{\lambda,i} / \sigma_{\lambda,i}^2 }{ \sum_i 1 /  \sigma_{\lambda,i}^2 },
\end{equation}

where $F$ is the stacked spectrum, $f_{\lambda,i}$ is the flux point at wavelength $\lambda$ of galaxy $i$ and $\sigma_{\lambda,i}$ the corresponding error. The sum is done over galaxies in any given cell.

The resulting error in each wavelength point of the stacked spectrum is simply

\begin{equation} \label{eq:stack_error}
\sigma^2_\lambda =  \frac{1}{ \sum_i 1 /  \sigma_{\lambda,i}^2}. 
\end{equation}

Note that by weighting the spectra within each stack based on signal-to-noise we are not modelling the total luminosity of the galaxies within the bin when we fit to the stacked spectrum. Instead our results should be interpreted as giving the history of {\em a weighted function} of the galaxies within that bin, although this weighting is actually close to a luminosity weighting, as spectral S/N is closely related to luminosity. 

The typical number of galaxies per stack is 1000, and we have a total of 124 stacks.

\subsection{VESPA and the fossil record}

We use VESPA to interpret the stacked spectrum of each cell in terms of a star formation and metallicity history. The algorithm is described in detail in \cite{TojeiroEtAl07}, and it has been applied to SDSS's LRG and Main Galaxy samples resulting in a queryable database, described in \cite{TojeiroEtAl09}. We refer the reader to these two papers for details but, for completeness, we present now a very brief summary of the method.

In short, VESPA inverts the equation $F_\lambda=\hat{F}_\lambda$, where $F_{\lambda}$ is the observed rest-frame flux, and the model flux is
\begin{equation}
\label{eq:vespa_problem}
  \hat{F}_{\lambda}(t_0) = \int_0^{t_0} f_{dust}(\tau_\lambda, t) 
  \psi(t) S_{\lambda}(t, Z)dt,
\end{equation}
to solve for the star formation rate $\psi(t)$ (solar masses formed per unit of time) and the age $t$ and metallicity $Z$ of the components of the stellar populations that each give luminosity per unit wavelength $S_{\lambda}(t,Z)$, per unit mass. The dependency of the metallicity on age is unconstrained, turning this into a non-linear problem. To get around this, we interpolate between the tabulated values of $Z$ given by the SPS models (see Section~\ref{sec:models}) giving a piecewise linear behaviour:
\begin{equation}
S'_\lambda(t,Z) = g(t) S_\lambda(t,Z_a) + \left[1-g(t)\right]S_\lambda(t,Z_b),
\end{equation}
where $S(t,Z_a)$ and $S(t,Z_b)$ are equivalent to $S_\lambda(t,Z)$ in Eq.~(\ref{eq:vespa_problem}), but at fixed metallicities $Z_{a}$ and $Z_{b}$, which bound the true $Z$.

Solving the problem then requires finding the correct metallicity range. One should not underestimate the complexity this implies - trying all possible combination of consecutive values of $Z_a$ and $Z_b$ in a grid of 16 age bins would lead to a total number of calculations of the order of $10^9$, which is unfeasible even in today's fast personal workstations. We work around this problem using an iterative approach, which we describe in \citet{TojeiroEtAl07}.

As for the treatment of dust, we follow the two-parameter dust model of \cite{CharlotFall00} in
which young stars are embebbed in their birth cloud up to a time
$t_{BC}$, when they break free into the inter-stellar medium (ISM):

\begin{equation}
f_{dust}(\tau_\lambda, t) = \left\{
\begin{array}{l}
f_{dust}(\tau^{ISM}_\lambda) f_{dust}(\tau^{BC}_\lambda), t \leq t_{BC}\\
f_{dust}(\tau^{ISM}_\lambda), t > t_{BC}\\
\end{array}
\right.
\end{equation}
where $\tau_\lambda^{ISM}$ is the optical depth of the ISM and
$\tau_\lambda^{BC}$ is the optical depth of the birth cloud. In previous runs of VESPA, we took $t_{BC}= 0.03$ Gyrs. However, in this run we have reduced resolution at the young end (see Section \ref{sec:age_grid} for a discussion on the revised age grid) and we do not resolve star-formation under $74$ Myr. We therefore take this boundary as $t_{BC}$, and note that removing this dust component altogether has minimal effect in our results and conclusions.

There is a variety of choices for the form of
$f_{dust}(\tau_\lambda)$. To model the dust in the ISM, we use the mixed slab model of
\cite{CharlotFall00} for low optical depths ($\tau_V \le 1$), for which
\begin{equation}
f_{dust}(\tau_\lambda) = \frac{1}{2\tau_\lambda}[1 + (\tau_\lambda -
1)\exp(-\tau_\lambda) - \tau_\lambda^2E_1(\tau_\lambda)]
\end{equation}
where $E_1$ is the exponential integral and
$\tau_\lambda$ is the optical depth of the slab. This model is known
to be less accurate for high dust values, and for optical depths
greater than one we take a uniform screening model with
\begin{equation}
f_{dust}(\tau_\lambda) = \exp(-\tau_\lambda).
\end{equation}
We only use the uniform screening model to model the dust in the birth
cloud and we use $\tau_\lambda =\tau_V(\lambda/5500\AA)^{-0.7}$ as our
extinction curve for both environments.

Recovering the star formation history, and its dust content, is solved by making assumptions about the form of $f_{dust}$ and $S_{\lambda}(t, Z)$ and by minimizing 
\begin{equation}
\chi^2 = \frac{ \sum_{\lambda} ( F_\lambda - \hat{F}_\lambda)^2}{\sigma_\lambda^2 }.
\end{equation}
Even though the problem has an analytical solution, a dataset perturbed by noise or that is otherwise deteriorated leads to instabilities in the matrix inversion and the recovered solutions can be entirely dominated by noise \citep{OcvirkEtAl06}. VESPA has a self-regularization mechanism which estimates how many independent parameters one should recover from a given a dataset that has been perturbed. The result is a parametrization which varies from galaxy to galaxy, depending on its SNR and wavelength coverage. 

It is the loss of resolution in lookback time, for low SNR spectra, that prompts the need for stacking in data-space. In other words, given the self-regularization mechanism and the inherent loss of - noisy! - information in cases of poor-quality data, it is {\em not} equivalent to stack in data or solution space. So, in general, the average of the star formation histories of the individual galaxies is not the same as the star formation history of the average spectrum. This becomes less and less true for increasingly better data of the individual objects, but in the case of the LRGs the data quality is sufficiently poor - especially at higher redshift - that stacking becomes necessary.

The physical quantities output by VESPA for each stacked spectrum are:
\begin{itemize}
\item star formation fraction in time interval $\Delta t_j$, $x_j$;
\item mass-weighted metallicity for mass formed in time interval $\Delta t_j$, $Z_j$;
\item dust attenuation for stars younger than 0.074 Gyrs, $\tau_{BC}$;
\item dust attenuation for all stars, $\tau_{ISM}$.
\end{itemize}


Given our choice for stacking method and the lack of overall normalisation, we cannot convert the mass fractions, $x_j$, into an absolute mass recovered in each time interval. This, however, does not pose a problem as we aim to compute a set of {\em colour} evolution tracks, for which an overall normalisation is not needed. For any given galaxy within a cell, the overall normalisation is given by its redshift and apparent magnitudes.

\subsection{The age grid of VESPA}\label{sec:age_grid}

In its original configuration, the VESPA age grid runs between 0.002 and 14 Gyr, in logarithmically spaced time bins. This, however, means that the width of the final bin is around 5 Gyr. This in turn is comparable to the stretch of lookback time that the LRG sample probes (roughly between z=0.5 and z=0.1) and we therefore have very little sensitivity to the exact formation epoch of LRGs. The original configuration was so designed because typical SDSS data for individual objects does not normally allow us to distinguish between populations that are separated in age by less than the bin widths.

The large SNR obtained from stacking, however, opens up the possibility to use a revised grid that has more sensitivity to older populations, at the expense of losing time resolution at the young end (the number of bins and overall structure must remain the same, for reasons constrained by the algorithm). We therefore construct an alternative age grid that we use to analyse the LRGs with VESPA. The boundaries of this new grid are: 0.002, 0.074, 0.177, 0.275, 0.425, 0.658, 1.02, 1.57, 2.44, 3.78, 5.84, 7.44, 8.24, 9.04, 10.28, 11.52 and 13.8 Gyr.
Whereas these seem ambitious, statistically the data should be able to differentiate between models in adjacent bins: the differences between models of adjacent ages (normalised to a common value at a given wavelength) are several times that of the statistical noise in the stacked data. In practice, however, systematic errors can dominate - see the next Section for details.

Finally note that given the high SNR of the stacked spectra, VESPA always runs to its highest time-resolution. 

\subsection{Error analysis}\label{sec:new_error}

VESPA was designed to deal with limitations coming from photon-noise. However, we are purposely putting ourselves in the limit where the photon-noise is negligible, and the quality of the fits is dominated by limitations in the modelling. We do not have a set of models that give statistically good fits in the limit of high-SNR and this has two potential effects: i) an error budget dominated by systematics in the modelling inaccuracies; and ii) the recovered parameters are biased. 


To deal with the model inaccuracies, we make the assumption that if the models (SSPs, dust and parametrization) could represent galaxies perfectly, then the recovered solution would match the stacked spectrum exactly. Under this assumption, the residuals from a fit in each cell can be interpreted as an overall error including statistical photon-noise errors and systematic model errors.

\begin{enumerate}
\item we run VESPA on the stacked spectrum and errors given by equations (\ref{eq:stack_error}), and calculate the residuals $r_\lambda = | \hat{F}_\lambda - F_\lambda|$;
\item we smooth $r_\lambda$ with a box-car filter to keep the large-scale variation but attenuate pixel-to-pixel fluctuations;
\item we take the smoothed $r_\lambda$ as the new error, $\sigma^{total}_\lambda$ and re-fit the stacked spectrum using this new error.
\end{enumerate}

\begin{figure}
\includegraphics[width=3.5in]{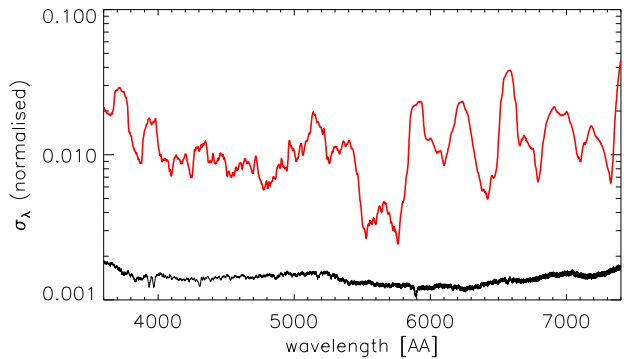}
\caption{Estimating a new error vector. Red: residuals smoothed with a box-car filter; black: photon-noise errors from the stack given by Eq.~(\ref{eq:stack_error}). The new estimated error larger than the stack error by around a factor of 10.}
\label{fig:new_noise}
\end{figure}

Fig.~\ref{fig:new_noise} shows a typical example of the residuals and the new estimated error. Therefore we allow extra freedom in the regions where the models cannot fit the data, by assuming that the residual between data and best-fit model gives an indication of the 1-sigma confidence region for systematic problems with the models, combined with the statistical error. As covariances will vary between different models, we assume this systematic is uncorrelated for flux values at different wavelengths, to avoid placing further biases on acceptable solutions. This estimate of the combined error replaces the instrumental statistical error in all of our fits. Our estimated systematic error will almost certainly be too small, because the best-fit parameters were chosen to match the data, and minimise this difference. Fundamentally, this is a problem that cannot be solved without better models. However the methodology outlined in this section should at least provide a approximation for the error and will be better than not allowing for this problem. 

We translate this error into an error bar on the recovered physical parameters by constructing $N$ random realisations of the best-fit spectrum, using the error calculated as per this section, and analysing these random realisations using VESPA. From this set of $N$ solutions we can estimate the variance and co-variance of all the parameters of interest. More details are given in \cite{TojeiroEtAl07,TojeiroEtAl09}. 

While these residuals provide a relative comparison of our current ability to model galaxies by comparing solutions obtained with different sets of models (SSPs, dust, etc), we must be careful not to interpret any variation in the recovered parameters as an absolute error bar - different models can be wrong in the same way. We analyse these differences in Section~\ref{sec:model_dependence}.

Finally, we should also note that the difference between models in adjacent bins (see Section \ref{sec:age_grid}) can be of the order of - or smaller than - this estimate of the systematic error. Ignoring the wavelength dependence, the average systematic error would correspond to roughly a 1 Gyr error on an 8 Gyr population.

\subsection{Smoothing the fossil record}

VESPA assumes that the star-formation rate (SFR) is constant within a time bin, and this gives results with discontinuities in SFR across bin boundaries. This is an obvious over-simplification of the problem, as any star formation history is likely to be continuous over an arbitrary set of boundaries, if sampled at high enough resolution. For the purposes of this paper, this over-simplification is problematic because a discontinuous SFR will result in a physically discontinuous colour evolution. 

To solve this problem, once VESPA has produce model parameters, we replace the top-hat bins of constant SFR with Gaussian distributions. The widths and heights are chosen to be such that the mass formed within each pair of the old sharp boundaries does not change by more than 0.3\%. The mass integrated over all ages is kept constant. These new Gaussian representations for each population (previously given by a single top-hat bin) are sampled at a much higher rate in lookback time, such that we can calculate a smooth colour evolution. Each new Gaussian bin has the same metallicity as the old corresponding histogram.

We do not expect this change in the SFR as a function of time to make a large difference in the interpretation of the spectra. This is because, by construction, the original boundaries are such that the spectral evolution between each boundary is not too significant.  The offset between the spectra calculated top-hat and Gaussian bins is either smaller or of the order of the statistical error shown in Fig.~\ref{fig:new_noise}. The flux obtained with the Gaussian bins is only used to compute the observed colours and their evolution with redshift, and the effect of this difference in the computed colours is negligible. 

\subsection{Evolving the observed spectrum}

Our goal is to compute the rest-frame spectrum of each cell (representing an ensemble of galaxies) at any point $t_e > t_g$, where $t_g$ is the lookback time of the cell in question. We compute
\begin{equation}\label{eq:F_te}
F_\lambda(t_e) = \sum_{t_k < t_e} m_k f_\lambda(t_k - t_e, Z_k),
\end{equation}
where $f(t,Z)$ is the flux of a population of age $t$ and metallicity $Z$, and $m_k = g(t_k) \Delta t_k$ is the mass formed at $t_k$, assuming a SFR given by $g(t)$. For the case where $t_e = t_g$ then we simply recover the observed spectrum, as shown in the section above. 

From $F_\lambda(t_e)$ we can easily compute any observed frame colours, by applying the relevant K+e corrections (fully known). As noted before, we cannot compute apparent magnitudes due to a lack of normalisation. However, for any individual galaxy the normalisation is set by the observed apparent magnitude and redshift, and $F_\lambda(t_e)$ gives then not only the colour, but also the luminosity evolution of that galaxy, assuming that the SFR in the galaxies follows the weighted average within the cell.
 
\subsection{Modelling dust}

Dust evolution does not leave a footprint in a galaxy's spectrum today, so it does not affect our modelling. However, in order to compare the colour tracks presented in Section~\ref{sec:colour_tracks} to observed populations of galaxies at different redshifts, we do need to treat the evolution of dust. Note that for the two-parameter dust model of \cite{CharlotFall00}, in addition to the inter-stellar dust component, young stars are allowed extra dust extinction to account for the effects of their birth cloud, which takes some time to dissipate, of the order $t_{BC}$. Both of these components are expected to evolve with redshift. We can envisage three possible solutions to this problem:
\begin{enumerate}
\item we can assume that the values we recover for $\tau_{BC}$ and $\tau_{ISM}$ remain valid for the entire history of the galaxy. I.e., when $t_e$ is such that a population becomes younger than $t_{BC}$ then we apply the value of $\tau_{BC}$ obtained at $t_g$, and we always apply the same value of $\tau_{ISM}$ to all population; and
\item we can estimate the evolution of $\tau_{BC}$ and $\tau_{ISM}$ from values observed for particular galaxies as a function of redshift, and apply these accordingly.
\item we can do the comparison in a dust-extinction corrected colour-colour or colour-redshift plane, in which case no dust evolution should be applied to the colour tracks.
\end{enumerate}

There are advantages and disadvantages to all methods. In the case of (i), we are over-simplifying the problem by ignoring a potential evolution with redshift, but we are keeping the information that relates to each individual cell. In the case of (ii) we are using a model for the evolution of dust, but we are assuming that the galaxies that we observe at low- and high-redshift are part of the same population. This latter assumption is exactly what we are looking to avoid by tapping into the fossil record. Option (iii) is probably the most robust, but requires modelling of galaxies at high redshifts. In this paper we plot the colour-tracks over {\em observed} colours, and for simplicity we apply the correction as given by (i) above. For other applications, colour evolution can be supplied with or without dust evolution.

\subsection{Summary of outputs}

It is worth summarising what the outputs of the analysis described in this section are. They can be broadly grouped into two types. On one hand we have {\em measured} quantities that relate to the position of the cell at a given redshift and colour, e.g. dust information ($\tau_{ISM}, \tau_{BC}$), mass-weighted metallicities or mass-weighted ages.  On another hand we have the colour-tracks as a function of redshift, which are inferred {\em evolved} quantities. All of these results have a dependence on the SSP synthesis codes. We describe the SSP models we use in the next section.

\section{Models} \label{sec:models}

VESPA models the spectrum of a galaxy as a superposition of simple stellar populations (SSPs) of different ages and metallicities. At the heart of this process lies the assumption that we know the spectral signature of different SSPs, and that we sample the full parameter space needed in order to model observed galaxies. 

There are three key steps that go into constructing an SSP. Firstly we need a description of a star's evolution given its mass and metallicity, in terms of observable parameters such as effective temperature or bolometric luminosity. This is traditionally given by the isochrones. Several groups have published sets of isochrones, which aim to model all stages of stellar evolution. Secondly, we must assume an initial mass function (IMF), which gives the correct weight as a function of mass, for stars formed in a single cloud of gas. Stars of different mass evolve with different time-scales, so by combining the isochrones and the IMF one can correctly populate a colour-magnitude diagram across the different stages of stellar evolution. The final step is to assign a spectrum to stars with different parameters, and this is done using the spectral libraries. Spectral libraries can be empirical or synthetic, with the former suffering from poor sampling of the parameter space of stellar evolution (dictated by the chemical enrichment of our own solar neighbourhood), and the latter suffering from deficiencies in our current ability to model stellar evolution. 

Crucially, the choices, treatments or calibrations involved in all of these three steps differ across different SPS codes, leading to unavoidable differences in the spectrum of SSPs of fixed age and metallicity. These differences are naturally centred around the least understood stages of stellar evolution, such as the thermally-pulsating asymptotic giant branch (TP-AGB) phase, the supergiant phase, or horizontal branch stars, and the datasets used for internal calibrations. The first efforts to accurately estimate the effects of the Horizontal Branch, Red Giant and TP-AGB branch on SSPs came from \cite{JimenezEtAl98, Maraston98, JimenezEtAl04, Maraston05}. \cite{JimenezEtAl98, JimenezEtAl04} built an analytic model to describe this stages of stellar evolution paying careful attention to use observations of individual and resolved stars to calibrate the stellar evolution models (see \citealt{JimenezEtAl95}). They pointed out how overestimations of the TP-AGB and lack of morphological description of the HB compromised seriously conclusions drawn from SSPs. They were, also, the first ones to emphasize the importance of stellar interior tracks being computed self-consistently with the stellar atmospheres models \citep{JimenezEtAl98, JimenezEtAl04}. Several authors have since followed this approach to estimate the broad-band SSPs (see e.g. \citealt{ConroyEtAl09} among others).

 \citet{ConroyEtAl09}, for example, look at the impact of these choices and assumptions on the recovered physical properties of galaxies, using broadband colours, with the aim of estimating the systematic error that arises from the limitations of the models. Here we do not take such a systematic approach, but we do conduct our analysis using three sets of popular SPS codes and investigate the resulting differences. The choices of IMF, stellar libraries or method for computing the stellar energetics cannot always be matched across different SPS codes. In each case we take the combination of these that best suits our application. We briefly describe the three sets of models next.

\subsection{BC03}

With the BC03 models \citep{BruzualEtCharlot03} we adopt a Chabrier initial mass
function \citep{Chabrier03} and Padova 1994 evolutionary tracks
\citep{AlongiEtAl93, BressanEtAl93, FagottoEtAl94a,
FagottoeEtAl94b, GirardiEtAl96}.  We use the BC03 models with the empirical STELIB library \citep{LeBorgneEtAl03} in the optical (3200\AA\  to 9500\AA ) and the theoretical BaSeL 3.1 stellar library \citep{LejeuneEtAl97, LejeuneEtAl98, WesteraEtAl02} to either side of that range. 

\subsection{M10}

The M10 models (Maraston \& Stromback , submitted) are the new, high-resolution version of the M05 models \citep{Maraston05}. The energetics and stellar evolution calculations are unchanged from the models in \citet{Maraston98} and M05, so the post main-sequence continues to be treated with the fuel-consumption theorem (\citealt{RenziniAndBuzzoni86, Buzzoni89, Maraston98}), leading to a more robust treatment of advanced stages of stellar evolution, such as the TP-AGB phase. The M10 models and publication focus not only on delivering a high-resolution version of the M05 models, but also on exploiting differences arising from stellar libraries and their calibration. Therefore, models are provided in a range of stellar libraries, each with different coverage in age, metallicity and wavelength. Here we choose the MILES library \citep{SanchezBlazquezEtal06}, as it provides the best coverage in age and metallicity. The wavelength coverage was extended to the UV following the method in \citet{MarastonEtAl09a}. We adopt a Kroupa IMF \citep{Kroupa01}.

\subsection{FSPS}

The Flexible Stellar Population Synthesis (FSPS) code \citep{ConroyEtAl09, ConroyAndGunn09} takes a novel approach to the computation of SSP spectra. By parametrizing uncertain stages in stellar evolution and allowing these parameters to change freely, the authors attempt to both quantify our ignorance about stellar evolution in terms of derived galaxy properties, but also to calibrate their models by finding the combination of parameters that best describe certain observations. Furthermore, models are provided with a choice of isochrones and stellar library. Here we chose the latest Padova evolutionary tracks calculations \citep{MarigoEtAl07, MarigoEtAl08}, and the MILES stellar library. An UV extension was obtained by using the theoretical BaSeL spectral library. In spite of the freedom provided by these models, here we use them simply at their default settings and with the combination of parameters that describe the TP-AGB phase that were found to best fit star cluster data by \citet{ConroyAndGunn09}. The opportunity remains to explore further the flexibility of these models - we leave that for a follow up paper. We use a \citet{Chabrier03} IMF. 

\subsection{$\alpha$ - enhancement}\label{sec:known_limitations}

The metallicities implicit in the SPS models and quoted throughout this paper refer to [Fe/H] abundances. However, it is now observationally well established that ETGs and bulges have a larger abundance of the so-called $\alpha$-elements (O, Ne, Mg, Si, S, Ca, Ti) than that predicted by SPS models, given the measured Fe abundances (e.g \citealt{WortheyEtAl92, DaviesEtAl93, FisherEtAl95, LonghettiEtAl00, MarastonEtAl03, ThomasEtAl03, ThomasEtAl05} and references therein). 

$\alpha-$elements are mostly produced through the collision of $\alpha$-particle nuclei in type-II supernovae of very massive stars, whereas Fe-peak elements are associated with type-Ia supernovae, typically associated with older and lower mass stars (e.g. \citealt{WoosleyWeaver95}). Given the different delay times for the two types of stellar explosions, different [$\alpha$/Fe] abundances have therefore been interpreted as a signature for the length of time over which the star formation happened (e.g. \citealt{PagelEtAl95, ThomasEtAl05}). The mismatch with the Fe abundances predicted by SPS models therefore occurs because empirical stellar libraries are primarily constructed by stars in the solar neighbourhood, which has a different chemical enrichment and star formation history than that which is typical of bulges and massive elliptical galaxies. 

Whereas some effort has gone into calibrating certain spectral indices for different $\alpha$-element abundances (e.g \citealt{ThomasEtAl03}), this has not yet been done for the full spectrum. Different chemical abundances are therefore a known limitation of current SPS models based on empirical libraries, if one wants to use a full-spectral fitting technique. 
Theoretical libraries however, can provide some insight on how different chemical abundances ratios affect galactic spectra (e.g. \citealt{CoelhoEtAl07}). For the moment, such discrepancies (surely to be systematic when looking at a population such as the LRGs) are automatically bundled with other unknown systematic errors in Section~\ref{sec:new_error}. 

\section{Results} \label{sec:results}

The main product of this work is a set of colour evolutionary tracks as a function of redshift and colour. In conjunction with high redshift data to match to, these can be used to solve some galaxy evolution cases, and we will consider such an analysis in a companion paper. In this paper we now explore some of the most immediate results, such as dependence of dust, age and metallicity on colour and redshift. 

The dependence on the underlying SSP modelling is investigated in Section~\ref{sec:model_dependence}. In this section, for clarity, we will often show results by choosing one of the three sets of SPS sets of models, and we append the matching plots using the other two sets at the end of this paper. Our choice of SSP model in the main text in each case should not be interpreted as supporting one over the others. When practical, we will show results from all three sets of models alongside each other. 

We start by showing some a typical fit and residuals.

\subsection{Typical fit and residuals}\label{sec:fits}

\begin{figure}
  \begin{center}
   \includegraphics[width=3.2in]{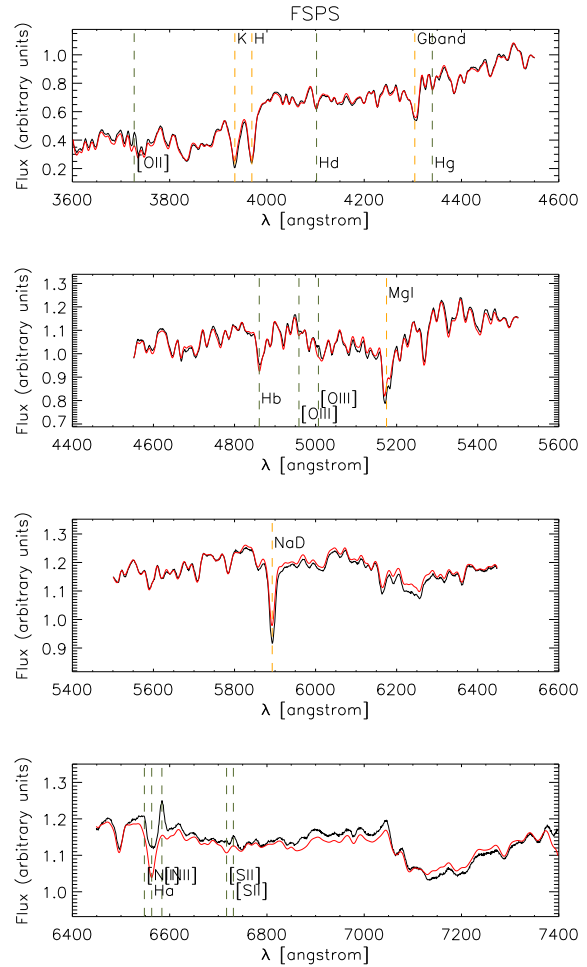}
    \caption{A typical fit, using the FSPS models (equivalent plots, for Mastro and BC03 models can be found in Appendix~\ref{sec:appendix_fits}). The black line is the data, and the red line the best-fitting model. The vertical yellow and green dashed lines guide the eye by showing some common emission and absorption features. Although plotted, some regions of the spectrum were excluded from the fit - see Fig.~\ref{fig:test_Residuals}}.
    \label{fig:test_fit}
  \end{center}
\end{figure}

\begin{figure}
  \begin{center}
   \includegraphics[width=3.2in]{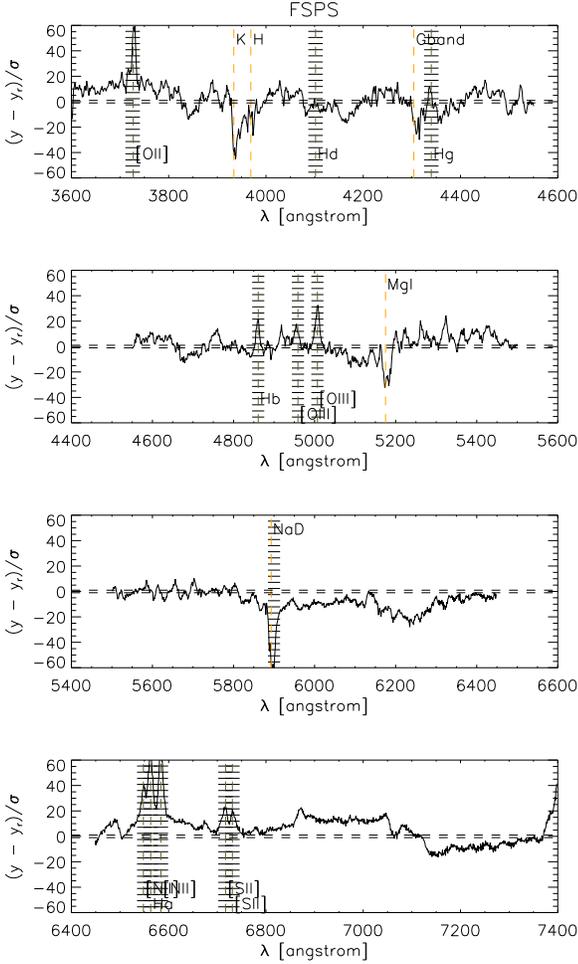}
    \caption{The residuals in units of the stack noise, for the fit showed in Fig.~\ref{fig:test_fit}. The vertical dashed lines guide the eye by showing some common emission and absorption features. The regions shaded by horizontal lines were excluded from the fit.}
    \label{fig:test_Residuals}
  \end{center}
\end{figure}

In Fig.~\ref{fig:test_fit} we show a fit for one of the stacks, and in Fig.~\ref{fig:test_Residuals} the corresponding residuals in units of the stack noise, as given by Eq.~(\ref{eq:stack_error}). This fit was obtained using the FSPS models (results for the same stack, with Mastro and BC03 models can be found in Appendix A). In this case we can see a visually good match on the blue end, and an increasing tension between data and models towards the red. The detail of the residuals plot reveals that there are significant departures from the data at all wavelengths, but more notoriously blueward of 3800\AA\ and redwards of 5800\AA\ .

Fig.~\ref{fig:test_Residuals} also shows regions of the spectrum that were excluded from fits that give the results presented in this section. Inclusion or exclusion of these regions does not affect the results in this section in any significant way.

\subsection{Measured quantities}\label{sec:measured_quantities}

In this section we show how ages, dust and metallicity vary with colour, luminosity and redshift in our LRG samples. The luminosity ranges refer to those shown in Fig.~\ref{fig:Mr_limits}.

When we show or quote errors in this Section, they refer to variations in colour, for a given redshift bin and luminosity slice.These variations are typically larger than the systematic error obtained as per Section \ref{sec:new_error}, which is likely to be an underestimation of the true systematic error. The effect of the statistical error on the recovered solutions is negligible.

\begin{figure*}
  \begin{center}
   \includegraphics[width=5.2in]{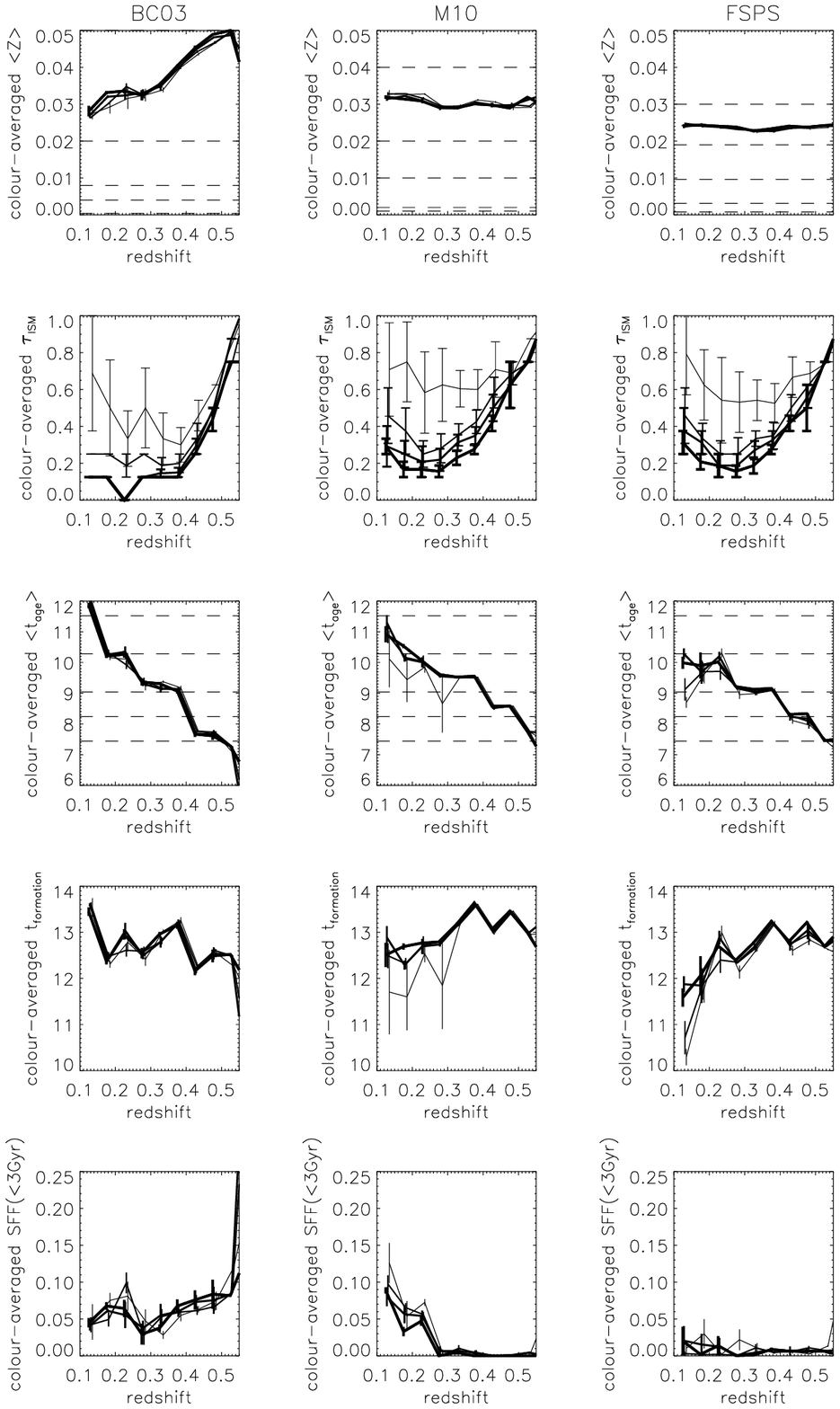}
    \caption{The evolution of mass-weighted metallicity (first row), dust extinction (second row), mass-weighted age (third row), formation epoch (forth row) and recent star-formation fraction (fifth row) with redshift for stacks of different luminosities. The thickness of the line in all plots represents the luminosity of the galaxies in each stack (thinner for the faintest, and thickest for the brightest). At each redshift, the quantities are averaged over $r-i$ colour. The horizontal dashed lines in the first row represent the metallicities provided by each set of models. Extrapolation beyond the extreme values is not permitted. The horizontal dashed lines in the 3rd row show the age boundaries used by VESPA. The error bars are the error on the mean over colour, and are representative of the variation with colour in a given redshift bin. See text for more details.}
    \label{fig:alltrends_allmodels}
  \end{center}
\end{figure*}

A summary of the results in this section can be found in Fig.~\ref{fig:alltrends_allmodels}.

\subsubsection{The ages of LRGs}  \label{sec:BB_prior}

The fiducial model is that LRGs formed mostly in one short epoch of star-formation at high-redshift. Therefore, the age of the oldest stars in LRGs of different redshifts should be consistent with a single epoch of formation. This is in principle testable and our revised age grid is sufficiently fine to test this hypothesis (see Section~\ref{sec:age_grid}).

In contrast, we find best-fit solutions given by the models that are systematically older than the age of the Universe \citep{KomatsuEtAl09}. Moreover, we found this behaviour with all three SPS models. 
We find that the best-fit solution is still too old in many cases, even after adjusting the error in the flux according to the method in Section~\ref{sec:new_error}
This suggests that the behaviour is systematic rather than statistically random. We therefore decide to impose a strong prior on possible solutions, such that star-formation cannot happen in a bin whose youngest boundary has an age older than the age of Universe. All results from hereon have this prior with the age of Universe set to be 13.8 Gyrs at z=0. The best-fitting solutions obtained with this prior have a formally worse $\chi^2$ than those without the prior, as expected. There is important information in the residuals of the solutions obtained with and without this prior - one is clearly wrong, but provides a better fit. The difference between the two sets of residuals should highlight which spectral features yield this difference. We leave this analysis to a companion paper.

\begin{figure*}
  \begin{center}
   \includegraphics[width=3.5in, angle=90]{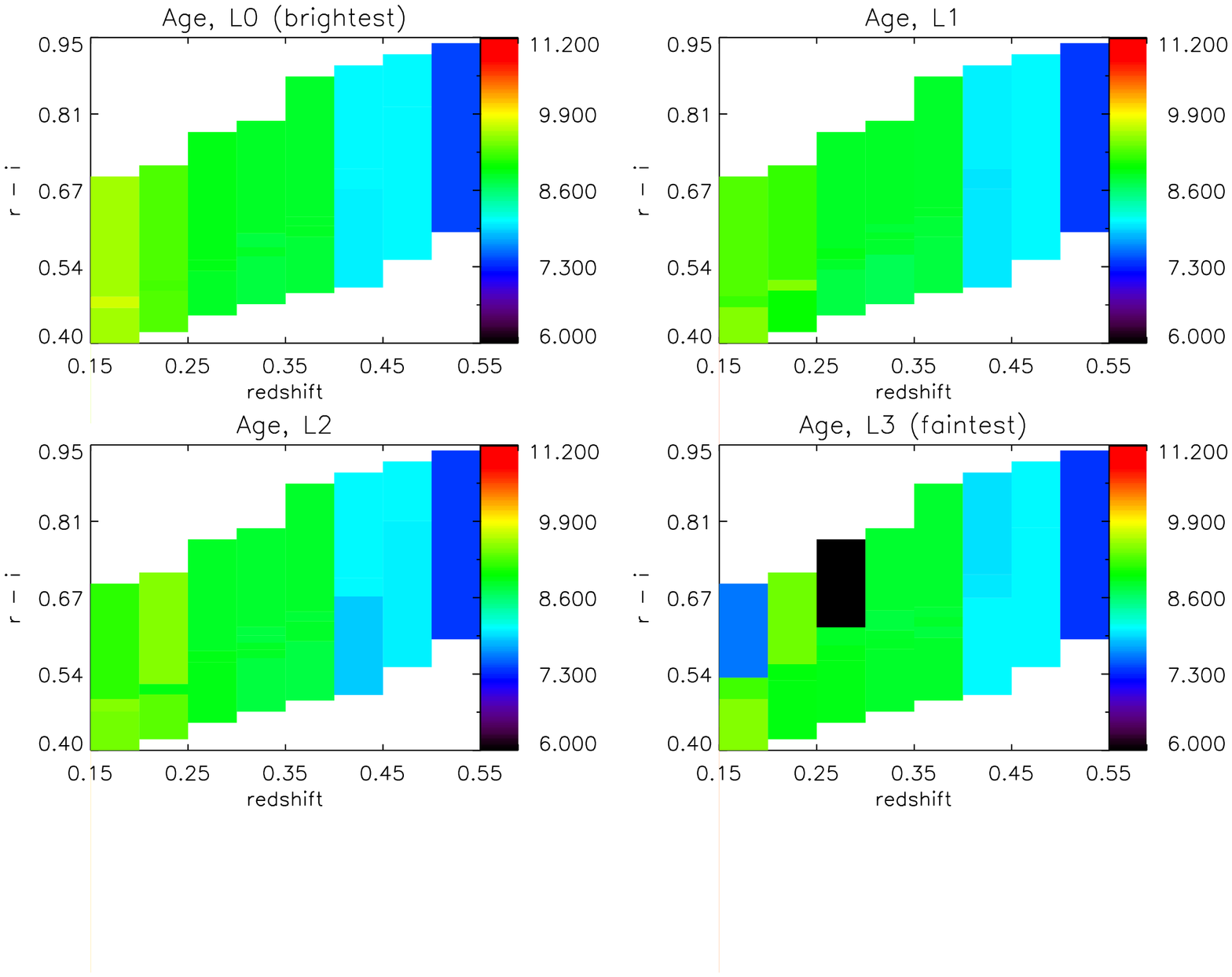}
    \caption{Mass-weighted age in Gyr (see Eq.~\ref{eq:mass_weighted_age}) for galaxies of different colour, redshift and rest-frame $r-$band luminosity, analysed with the M10 models. The luminosity ranges change with redshift, and are shown in Fig.~\ref{fig:Mr_limits}. We only show data for regions of parameter space with a sufficient number density of galaxies. In practice, there are a few galaxies (less than 20 per redshift bin) outside of the coloured areas, but these have an insignificant impact on the recovered solution. In Fig.~\ref{fig:alltrends_allmodels} (3rd row) we average over colour to show the trend with redshift for each luminosity slice.}
    \label{fig:ages_allhists}
  \end{center}
\end{figure*}

Fig.~\ref{fig:ages_allhists} shows the dependence of the rest-frame mass-weighted age of each stack of galaxies, as a function of colour and redshift for four luminosity slices, and calculated using the M10 models (identical plots obtained with BC03 and FSPS are given in Appendix~\ref{sec:appendix_2dhists}). We calculate a mass-weighted age as

\begin{equation}\label{eq:mass_weighted_age}
\langle t_{age} \rangle = \frac{\sum_{i} t_i x_i}{\sum_{i} x_i}
\end{equation}

where we take $t_i$ as being the mean age of bin $i$. We see no strong evidence for a dependence of $\langle t_{age} \rangle$ on colour, and we see the expected ageing of galaxies towards lower redshifts (largely dictated by our prior on the age of the Universe).

In Fig.~\ref{fig:alltrends_allmodels} (3rd row) we collapse the histogram in colour, and show the evolution in redshift obtained with different SPS models. The error bars show the error on the mean over colour, and so are representative of the variation with colour, in a given redshift bin. In the 4th row of Fig.~\ref{fig:alltrends_allmodels} we add the lookback time to the redshift of each stack, and show the formation time in the Earth-frame. 

\subsubsection{Recent to intermediate star formation}

\begin{figure*}
  \begin{center}
   \includegraphics[width=3.5in, angle=90]{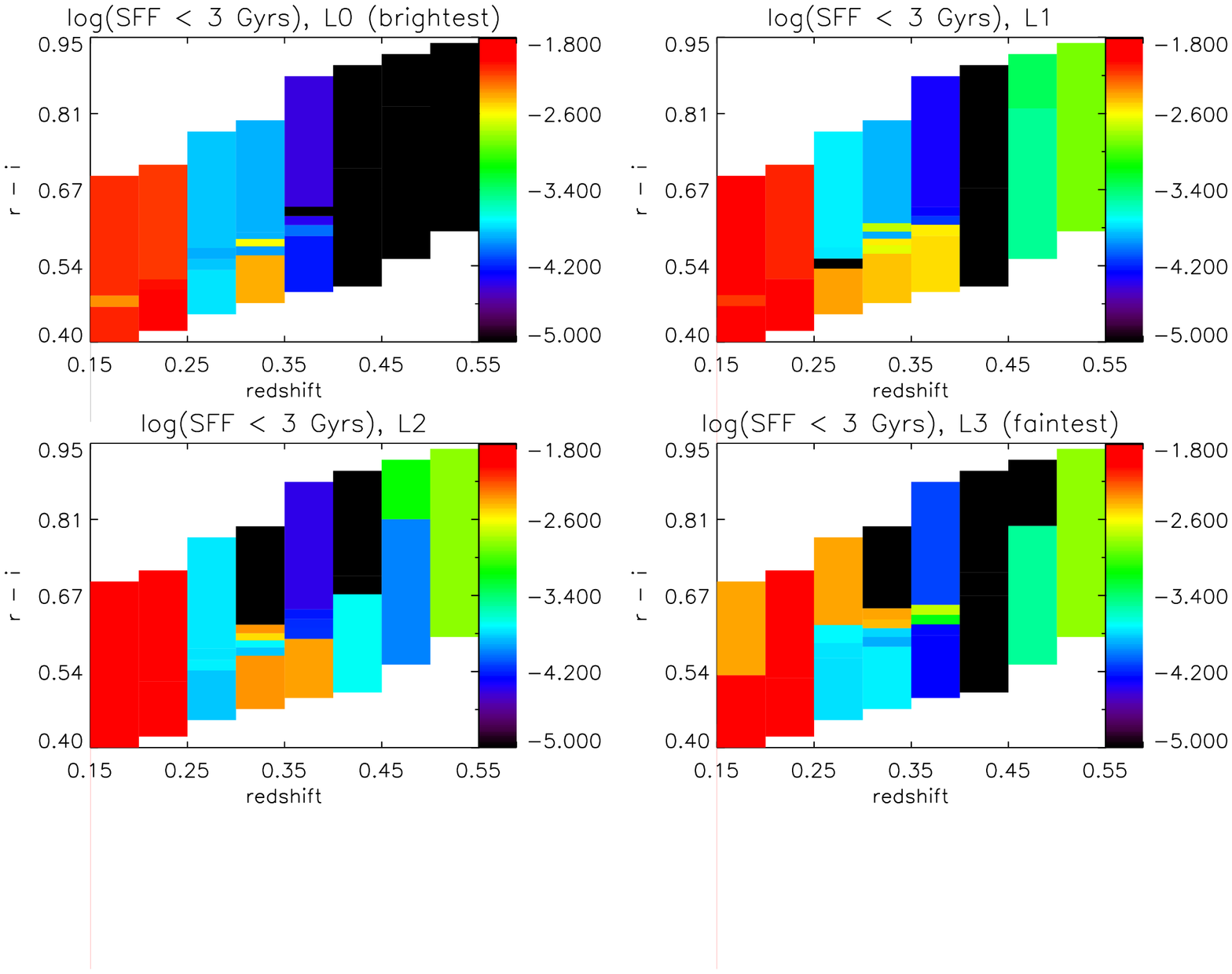}
    \caption{The logarithm of the fraction of star formation (by mass), recovered in bins up to 3.8 Gyr for galaxies of different colour, redshift and rest-frame $r-$band luminosity, analysed with the M10 models. The luminosity ranges change with redshift, and are shown in Fig.~\ref{fig:Mr_limits}. We only show data for regions of parameter space with a sufficient number density of galaxies. In practice, there are a few galaxies (less than 20 per redshift bin) outside of the coloured areas, but these have an insignificant impact on the recovered solution. In Fig.~\ref{fig:alltrends_allmodels} (5th row) we average over colour to show the trend with redshift for each luminosity slice.}
    \label{fig:RSF_2dhist_M10}
  \end{center}
\end{figure*}

Collapsing the full SFH into a mass-weighted age, however, can mask interesting information such as the amount of recent star formation. In Fig.~\ref{fig:RSF_2dhist_M10} and, collapsed in colour, in the 5th panel of Fig.~\ref{fig:alltrends_allmodels}, we show the fraction of star-formation, by mass, recovered in bins up to 3.8 Gyr in the rest-frame of each stack. A summary is given in Table \ref{tab:SFF}, where we split this range further in intermediate and young ages.

Although a fixed age in the rest frame is equivalent to a different fraction of the galaxies' ages as we go back in redshift, 3.8 Gyr serves as a good split between stars that formed in the oldest, fiducial burst, and anything that may have followed. We discuss these results in Section~\ref{sec:model_dependence}.

\begin{table*}
\begin{tabular}{|c|c|c|c|c|c|c|c|}
\hline \hline
\multirow{3}{*}{}  		   & 				&\multicolumn{2}{|c|}{$0.15<z<0.35$} & \multicolumn{2}{|c|}{$0.35<z<0.5$}  \\  
			SPS model & Luminosity range & Young & Intermediate                         & Young & Intermediate \\ 
                               		  & 				& SFF $< 0.7$ Gyr & SFF $0.7$ - $3.8$ Gyr  & SFF $< 0.7$ Gyr & SFF $0.7$ - $3.8$ Gyr \\ \hline \hline
                              
\multirow{4}{*}{BC03} &L0 (brightest)& 0.0018$\pm$0.00029 & 0.047$\pm$0.0085 & 0.0068$\pm$0.00024 & 0.068$\pm$0.0068\\
				   &L1 & 0.0019$\pm$0.00028 & 0.048$\pm$0.011 & 0.013$\pm$0.00028 & 0.126$\pm$0.0057 \\
				   & L2 & 0.0019$\pm$0.00033 & 0.054$\pm$0.011 & 0.010$\pm$0.00024 & 0.114$\pm$0.0023 \\
				   &L3 (faintest)& 0.0013$\pm$0.00023& 0.056$\pm$0.015 & 0.013$\pm$0.00029 & 0.0776$\pm$0.0040 \\ \hline
				   
\multirow{4}{*}{M10} &L0 (brightest) & 0.0018$\pm$0.00065 & 0.036$\pm$0.0033 & 0.00014$\pm$4e-06 & 0$\pm$0 \\
			          &L1 & 0.0019$\pm$0.00069 & 0.043$\pm$0.0071 & 0.00015$\pm$3e-06 & 0.00280$\pm$7e-05 \\
				& L2 & 0.0015$\pm$0.00057 & 0.046$\pm$0.0051 & 0.00044$\pm$1-06 & 0.0038$\pm$0.0001 \\
				& L3 (faintest) & 0.00311$\pm$0.0022 & 0.056$\pm$0.0098 & 0.0009$\pm$2e-06 & 0.018$\pm$0.0004 \\ \hline

\multirow{4}{*}{FSPS} &L0 (brightest)& 0.0068$\pm$0.0066 & 0.0017$\pm$0.0007 & 0$\pm$0& 0.0046$\pm$0.0013 \\
				  & L1& 0.0096$\pm$0.0094 & 0.0020$\pm$0.0002 & 0$\pm$0 & 0.0063$\pm$0.0015 \\
				 & L2 & 0.0022$\pm$0.0020 & 0.0025$\pm$0.0007 & 0$\pm$0 & 0.0048$\pm$0.00076 \\
				  & L3 (faintest)& 0.013$\pm$0.0086 & 0.0076$\pm$1e-05 & 0$\pm$0 & 0.026$\pm$0.00078 \\ \hline \hline
\end{tabular}
\caption{Summary of the recovered star-formation fractions at young and intermediate ages, averaged over two redshift ranges, for all luminosity slices and the three SPS models we consider. As in Fig.~\ref{fig:alltrends_allmodels}, the errors come from the variation with colour at a given redshift bin.}
\label{tab:SFF}
\end{table*}

\subsubsection{Metallicity}

\begin{figure*}
  \begin{center}
   \includegraphics[width=3.5in, angle=90]{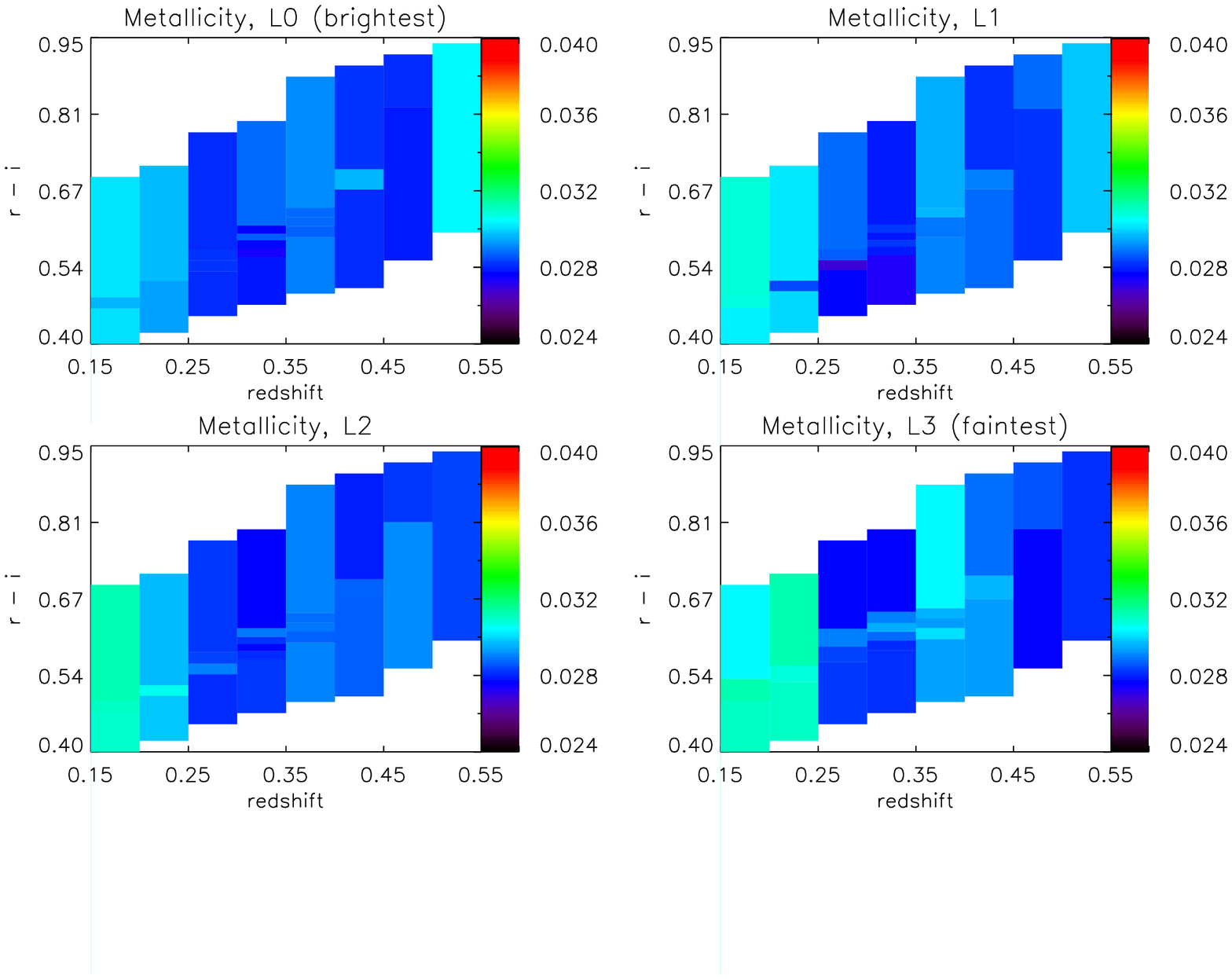}
    \caption{Mass-weighted metallicity (see Eq.~\ref{eq:mass_weighted_metallicity}) for galaxies of different colour, redshift and rest-frame $r-$band luminosity, analysed with the M10 models. The luminosity ranges change with redshift, and are shown in Fig.~\ref{fig:Mr_limits}. We only show data for regions of parameter space with a sufficient number density of galaxies. In practice, there are a few galaxies (less than 20 per redshift bin) outside of the coloured areas, but these have an insignificant impact on the recovered solution. Fig.~\ref{fig:alltrends_allmodels} averages over colour to show the trend with redshift for each luminosity slice.}
    \label{fig:metallicity_allhists}
  \end{center}
\end{figure*}

Fig.~\ref{fig:metallicity_allhists} shows the mass-weighted metallicity as a function of $r-i$ colour and redshift, for four luminosity slices, and calculated using the Mastro models (identical plots obtained with BC03 and FSPS are given in Appendix A). For each cell, we calculate the mass-weighted metallicity using the full star-formation history as

\begin{equation}\label{eq:mass_weighted_metallicity}
\langle Z \rangle = \frac{\sum_{i} Z_i x_i}{\sum_{i} x_i}.
\end{equation}

To make any possible trend with redshift more clear, we average over colour and show the resulting redshift dependence in Fig.~\ref{fig:alltrends_allmodels} (first row). The horizontal dashed lines show the metallicities provided by the models - note that different sets of models sample the metallicity space in different ways. We discuss the different behaviour of each set of models in Section~\ref{sec:model_dependence}. As in the previous section, the error bars are show the error on the mean over colour.

\subsubsection{Dust}

\begin{figure*}
  \begin{center}
   \includegraphics[width=3.5in, angle=90]{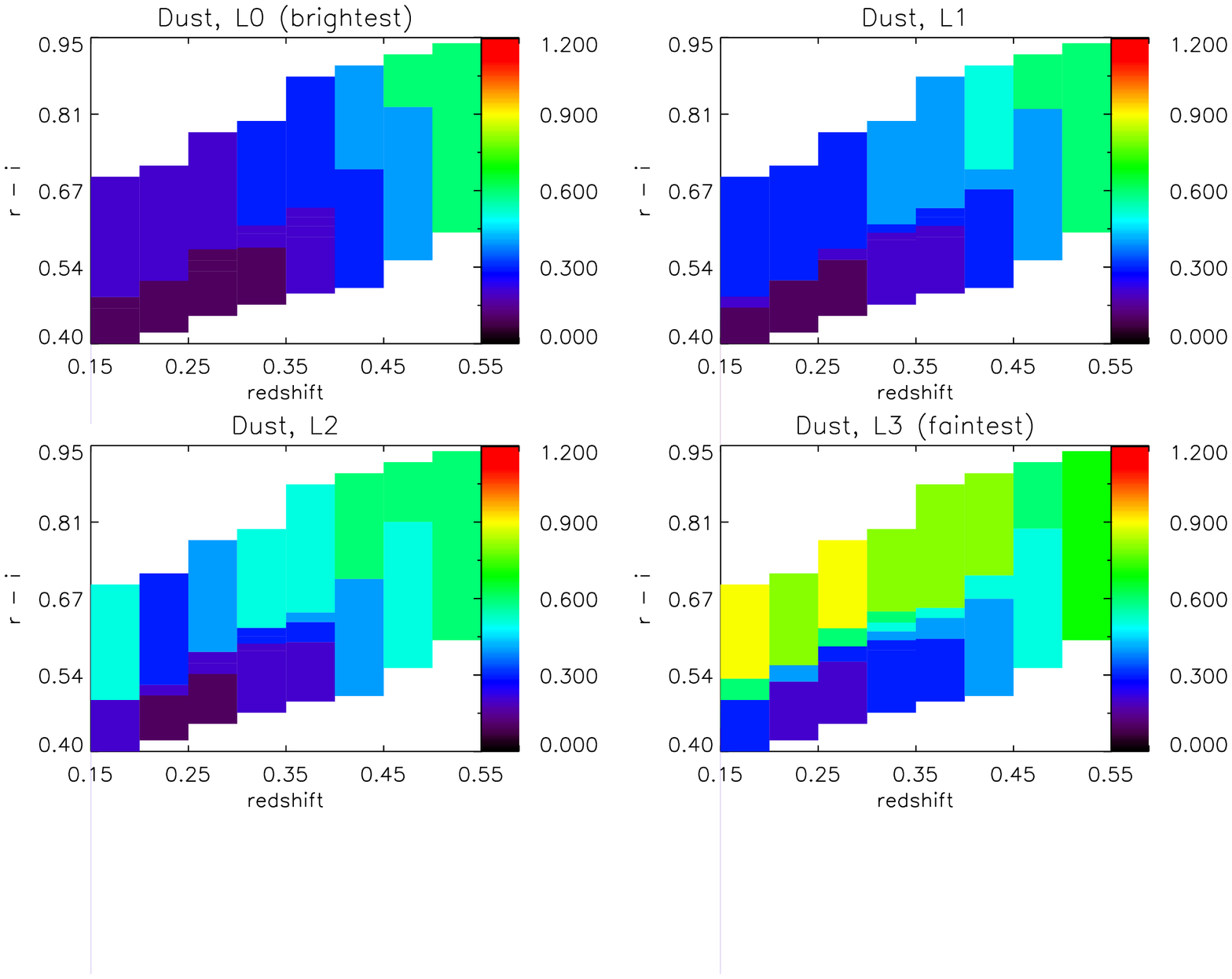}
    \caption{Inter-stellar dust absorption for galaxies of different colour, redshift and rest-frame $r-$band luminosity, analysed with the M10 models. The luminosity ranges change with redshift, and are shown in Fig.~\ref{fig:Mr_limits}. We only show data for regions of parameter space with a sufficient number density of galaxies. In practice, there are a few galaxies (less than 20 per redshift bin) outside of the coloured areas, but these have an insignificant impact on the recovered solution. Fig.~\ref{fig:alltrends_allmodels} averages over colour to show the trend with redshift for each luminosity slice.}
    \label{fig:dust_allhists}
  \end{center}
\end{figure*}

Fig.~\ref{fig:dust_allhists} shows the dust attenuation as a function of $r-i$ colour and redshift, for four luminosity slices. To make any possible trend with redshift more clear, we average over colour and show the resulting redshift dependence in Fig.~\ref{fig:alltrends_allmodels} (second row). The error bars are errors on the mean, and therefore give an indication of the scatter of dust with colour, for each redshift bin. We discuss these results in Section~\ref{sec:model_dependence}.

\subsection{Evolved quantities} \label{sec:evolved_quantities}

We can use the full star-formation history of each cell to estimate the colour evolution of a galaxy within that cell. This information can be coupled with the apparent magnitude and exact redshift of the galaxy to calculate its luminosity evolution. In this section we present a selection of these colour tracks for a sample of galaxies selected in colour, luminosity and redshift. As before, we present the results using only one set of SPS models in this section (FSPS), and provide the corresponding plots using the other two SPS models in Appendix~\ref{sec:appendix_colour_tracks}.

\subsubsection{Spectro and model magnitudes}

The fibre aperture in the SDSS has a fixed size of 3 arc-seconds. Apparent sizes of objects are often larger than this, leading to an unavoidable discrepancy between the magnitudes obtained from integrating a spectrum over a filter's response (the spectro magnitudes), and the best estimate of the photometric magnitudes (the cmodel magnitudes). We do our stacking using the latter, but we use the spectrum to obtain our star formation and metallicity histories, which in turn refer only to the 3 arc-second fibre. Therefore we need to be able to map one quantity to the other, and understand any possible biases in this relationship. The spectrophotometric calibration can also affect this mapping, and there is a known offset between the spectro and fibre magnitudes (obtained from the photometry, within the 3 arc-second aperture) since DR6 due to how this calibration is done \citep{Adelman-McCarthyEtAl08}. This offset, however, has an insignificant dependence on photometric band and it is not a problem for our colour selection.

\begin{figure}
  \begin{center}
   \includegraphics[width=2.7in]{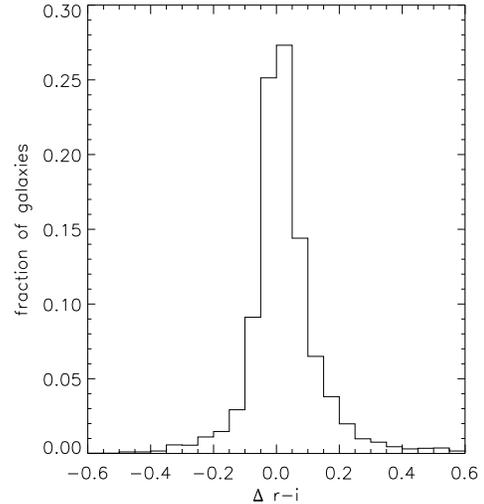}
    \caption{The distribution of $\Delta r-i$ (Eq.~\ref{eq:deltaC}) for a random sample of 10,000 LRGs.}
    \label{fig:deltaC}
  \end{center}
\end{figure}
\begin{figure}
  \begin{center}
   \includegraphics[width=2.7in]{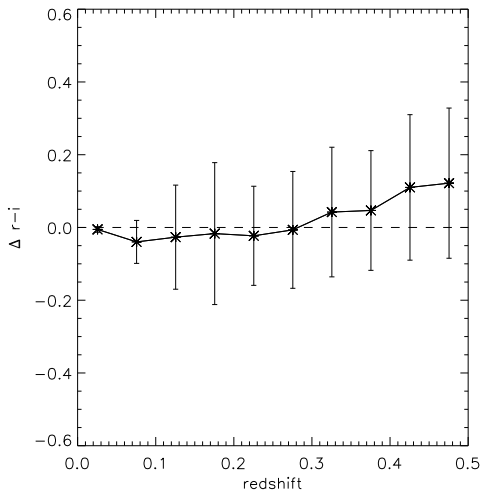}
    \caption{The mean and scatter of $\Delta r-i$ as a function of redshift. See text for an explanation of the upturn at high redshift. }
    \label{fig:DeltaC_redshift}
  \end{center}
\end{figure}

Fig.~\ref{fig:deltaC} shows the distribution of 
\begin{equation}\label{eq:deltaC}
  \Delta[r-i] = [r-i]_\mathrm{spectro} - [r-i]_\mathrm{cmodel},
\end{equation}
for a random sample of 10,000 LRGs. This scatter can be better understood by explicitly writing
\begin{equation}
  [r-i]_\mathrm{cmodel} = [r-i]_\mathrm{3''} + \delta_\mathrm{photo} 
  + [r-i]_{\mathrm{extra}}
\end{equation}
\begin{equation}
[r-i]_\mathrm{spectro} = [r-i]_\mathrm{3''} + \delta_\mathrm{spectro}, 
\end{equation}
where $[r-i]_{\mathrm{extra}}$ is the difference in colour due to a radial colour gradient extending outwith the 3 arc-second fibre, and $\delta_\mathrm{spectro}$ and $\delta_\mathrm{photo}$ are noise in the spectroscopic and photometric measurements, respectively. Assuming that $\delta_\mathrm{spectro}$ and $\delta_\mathrm{photo}$  are stochastic components, an offset from zero in $\Delta[r-i]$ would be indicative of a systematic behaviour in $[r-i]_{\mathrm{extra}}$. Fig.~\ref{fig:deltaC} shows no evidence for such an offset. Fig.~\ref{fig:DeltaC_redshift} shows $\Delta[r-i]$ as a function of redshift - there is an indication of a positive slope at high-redshift. A positive $\Delta[r-i]$ means a larger spectro colour, which in turn means a bluer colour inside the fibre, when compared to the full aperture. This is counter-intuitive for two reasons - firstly one would expect the proportion of light falling into the 3 arc-second fibre to go up with redshift, as apparent sizes get smaller; and secondly, one would expect the outer regions to be bluer in colour in comparison to the central regions. 

Note, however, that at high redshift we are intrinsically selecting increasingly redder galaxies (in cmodel colours) due to the LRG sample selection. These, in turn, are likely to have the largest positive values of $\delta_\mathrm{photo}$ and, in consequence, of $\Delta [r-i]$. There is an intrinsic Malquist-type selection bias in $\Delta [r-i]$ simply due to the fact that we select in $[r-i]_\mathrm{cmodel}$. The selection of galaxies for each cell will be biased in a similar way, but this bias is known as we can compute $\Delta [r-i]$ for each cell. 

Without independent photometry, it is hard to disentangle the redshift-dependence of this selection bias from a true dependence of $[r-i]_{\mathrm{extra}}$ in colour, or redshift. However, Figs.~\ref{fig:deltaC} and~\ref{fig:DeltaC_redshift} show no evidence of a detectable signal in $[r-i]_{\mathrm{extra}}$. Therefore, to translate from spectro colour to cmodel colour we simply apply a correction to the beginning of each track. We do this in all of the plots in the next section.

\subsubsection{Colour tracks} \label{sec:colour_tracks}

\begin{figure}
  \begin{center}
   \includegraphics[width=3.5in]{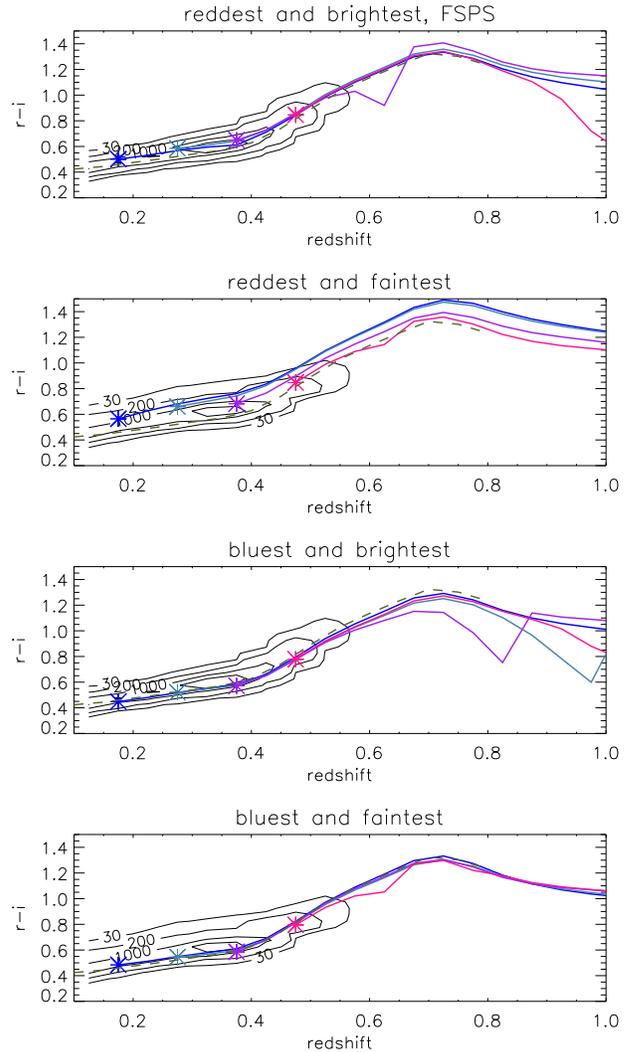}
    \caption{Predicted evolution of $r-i$ colour with redshift, calculated according to Eq.~(\ref{eq:F_te}), and the solutions presented in Section~\ref{sec:measured_quantities}. The different panels show examples at different combinations of colour and luminosity. The black contour lines show the number density of LRGs within the respective luminosity slice, and the coloured lines show the tracks from cells at four different redshifts: 0.175, 0.275, 0.375 and 0.475. The star shows the position of the cell the track relates to. The green dashed line shows the LRG model from \citet{MarastonEtAl09}, for comparison.}
    \label{fig:tracks_rmi}
  \end{center}
\end{figure}

\begin{figure}
  \begin{center}
   \includegraphics[width=3.5in]{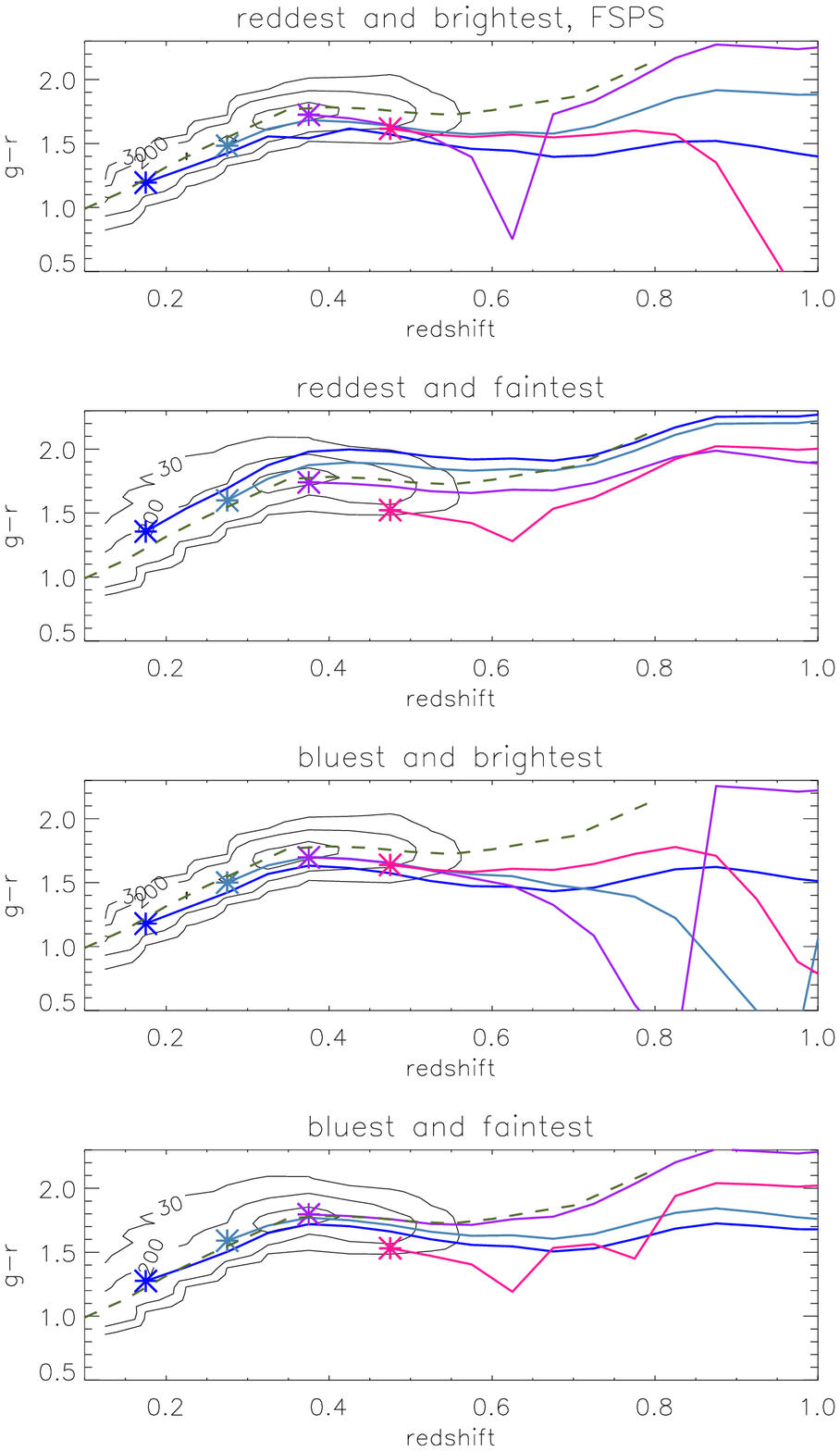}
    \caption{Predicted evolution of $g-r$ colour with redshift, calculated according to Eq.~(\ref{eq:F_te}), and the solutions presented in Section~\ref{sec:measured_quantities}. The different panels show examples at different combinations of colour and luminosity. The black contour lines show the number density of LRGs within the respective luminosity slice, and the coloured lines show the tracks from cells at four different redshifts: 0.175, 0.275, 0.375 and 0.475. The star shows the position of the cell the track relates to. The green dashed line shows the LRG model from \citet{MarastonEtAl09}, for comparison.}
    \label{fig:tracks_gmr}
  \end{center}
\end{figure}

\begin{figure}
  \begin{center}
   \includegraphics[width=3.5in]{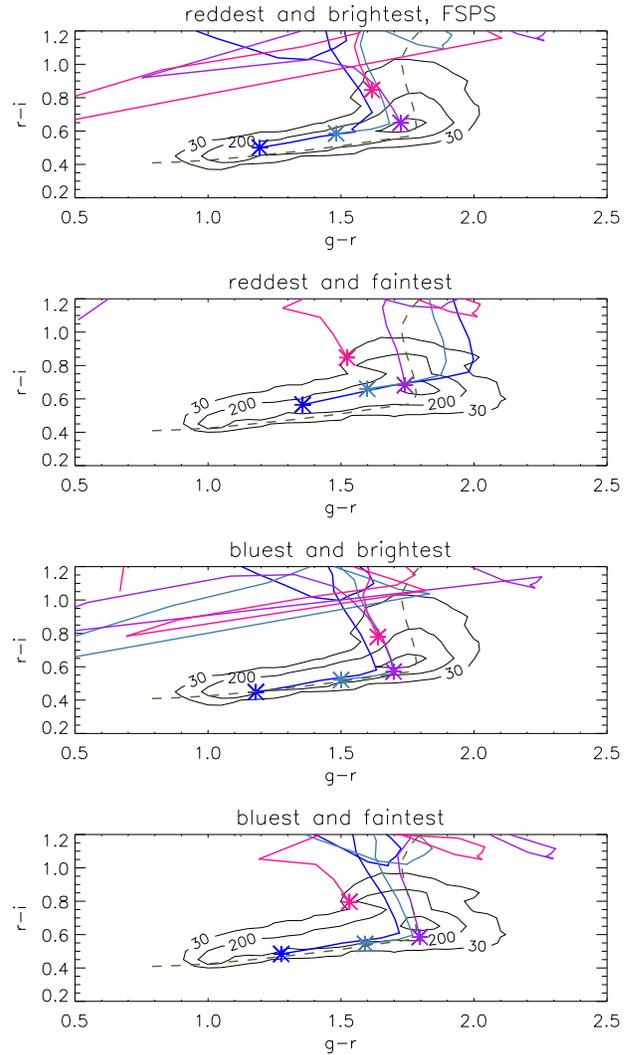}
    \caption{Predicted evolution of $r-i$ colour with $g-r$, calculated according to Eq.~(\ref{eq:F_te}), and the solutions presented in Section~\ref{sec:measured_quantities}. The different panels show examples at different combinations of colour and luminosity. The black contour lines show the number density of LRGs within the respective luminosity slice, and the coloured lines show the tracks from cells at four different redshifts: 0.175, 0.275, 0.375 and 0.475. The star shows the position of the cell the track relates to. The green dashed line shows the LRG model from \citet{MarastonEtAl09}, for comparison.}
    \label{fig:tracks_colour}
  \end{center}
\end{figure}

Figs.~\ref{fig:tracks_rmi} and~\ref{fig:tracks_gmr} shows the predicted evolution of $r-i$ and $g-r$ colour with redshift, for four combinations of colour and luminosity and obtained with the FSPS models. Fig.~\ref{fig:tracks_colour} shows this evolution in the $r-i$ vs $g-r$ plane, using the same models. In each case we over-plot with contour lines the number of {\em observed} LRGs for in the same luminosity range. The contours are wider for the fainter sample, but we cannot immediately tell whether this is due to an increased photometric error for fainter objects, or due to any intrinsic colour-luminosity relation. We explore these results and their model dependence in Section~\ref{sec:model_dependence}.

\section{Model dependence and interpretation}\label{sec:model_dependence}

The results presented in the previous section cannot be interpreted without addressing the fact that each set of models often gives a different result. In this paper we do not directly tackle the question of which set of models is the most correct, we limit ourselves to a very basic exploration of the residuals in Section~\ref{sec:goodness_of_fit}. 

Overall, we find agreement with the fiducial model that says that LRGs are dominated by old, metal rich stars. However, we find that some small departure of this model is needed to fit the stacked spectra, and that small order changes to the fiducial model depend on the SPS model used. Ultimately, we are interested in seeing how the stellar evolution of LRGs translates into colour tracks, and the potential implication for galaxy evolution and observational cosmology. We begin by analysing the in-situ results of Section~\ref{sec:measured_quantities} and then we discuss the effects on the colour tracks of Section~\ref{sec:evolved_quantities}.

\subsection{Measured quantities}

Fig.~\ref{fig:alltrends_allmodels} summarises the in-situ measurements presented in Section~\ref{sec:measured_quantities}. 

We see a different behaviour in the metallicity evolution when using different models, but all models consistently show LRGs to be metal-rich, as expected following the fiducial model. This difference is not a surprise - metallicity is known to be more model dependent than age (e.g. \citealt{PanterEtAl08,TojeiroEtAl09}), and even the simple fact that the models sample the metallicities at different intervals will have some effect.

The sharp increase in metallicity seen when using BC03 models is most likely due to the age-metallicity degeneracy. BC03 predicts a modest amount of  recent star-formation at low-redshift, but this fraction increases to higher redshifts (5th panel). The results is a mass-weighted formation time that gets younger with redshift, and an increasing metallicity. It is worth noting that the fraction of mass in young/intermediate stars is not generally uniformly distributed over the last 3 Gyr of the galaxies. Instead, VESPA generally assigns the mass to one or two age bins, and the particular bins with the most mass change from stack to stack. In the case of BC03 and M10, these are firmly at the intermediate ages regime. This is probably indicative of sporadic star formation episodes rather than a continuous star formation type of history, for any given stack. 

The FSPS and M10 models paint a scenario much closer to the fiducial model of passive stellar evolution, and a constant high-metallicity. However, the M10 models require a small amount of intermediate star formation at low redshift (a little over 10\% by mass for the faintest objects, and slightly less for brightest ones), which in turn drives down the mass-weighted ages of the faintest objects at low redshift. Note that this is different from the reason that drives down the mass-weighted ages of the faintest objects in FSPS - in this case it is the dominant, oldest populations that are slightly younger. This type of distinction is possible due to having time-resolved star formation histories, and it becomes crucially important if one aims to use the age-redshift relationship of LRGs to put tight constraints on $H(z)$ (e.g. \citealt{ JimenezEtAl03, SimonEtAl05, CrawfordEtAl10, SternEtAl10}).

Aside from the dust results, we see little coherent dependence on luminosity, in either age, metallicity or amount of recent star formation. Work on ETGs has consistently shown that fainter objects tend to have younger stars on average (e.g. \citealt{ThomasEtAl05, CarsonEtAl10, ZhuEtAl10} ), but the reason why we do not see clearly in this study is easily understood. The range in luminosity of our LRG sample is much smaller than the traditional ETG samples, which tend to go down to fainter magnitudes, leaving us with a small lever-arm in luminosity with which to constrain this type of relationship. Furthermore, as the LRG sample is neither magnitude nor volume limited, any trend with luminosity over redshift is difficult to interpret. We do see a tendency for fainter objects to be younger, in both the FSPS and M10 cases (although for different reasons, as explained in the previous paragraph), but this is limited to $z \lesssim 0.3$. This is explained both by the lack of faint galaxies in the sample at larger redshifts, and by the larger range in magnitudes in the sample at low redshifts.

Dust extinction results are relatively consistent across different models. This is likely to be due to the fact that the dust modelling is common in all three cases, and that the large-scale changes due to the presence of dust are independent enough of large-scale changes due to age or metallicity. We consistently see a very clear separation in luminosity - fainter objects have more dust extinction than bright objects - and an increase towards high-redshift. In the case of Mastro and FSPS, we also see evidence for an upturn at low redshifts, suggesting that perhaps there is some contamination at his end by dustier galaxies (with some recent star formation in both cases, although up to 10 times more, in fractional mass, in the case of the Mastro models). Additionally, dust extinction is the only of our measurements to show a clear trend with colour - redder cells tend to have larger extinction, across all luminosities. 

\subsection{Evolved quantities}

Our particular interest in this paper is to see how the solutions discussed in the previous section translate into the colour evolution of LRGs of different colour, luminosity and redshift. Inevitably, the model dependence observed in the previous section will manifest itself in the colour tracks.

The most obvious is the effect of recent star formation. An episode of star formation results in both a short and abrupt change in the colour tracks, and a reddening of the colours to higher redshift. The magnitude of these two effects depends on the size (in mass), age and duration of the episode of star formation. Metallicity and dust induce less abrupt changes in the colour evolution. 

The interpretation of the results from the different sets of models is rather different. In the case of FSPS, with negligible amounts of recent star formation and a constant metal-rich population, the colour tracks follow the locus of observed objects very closely. As the selection box was designed around a passively evolving model, this is no less the expected behaviour. Furthermore, we see little dependence on colour or luminosity, suggesting that the differences noted in the previous section (namely dust and age of oldest population) have a small effect on the colour evolution.

It is a different matter in the case of the M10 or BC03 models. With BC03, the substantial and ever-present (in redshift, colour and luminosity) level of recent star formation means that star formation events constantly push the colour tracks outside the contours. Depending on the duration of the burst (and we have limited resolution with our grid) then an object remains outside the selection for box for a given amount of time, until the excessively blue light of the young stars subsides enough to make the object red once again. In principle, there is nothing wrong with this scenario, and it follows that if LRGs go through small and sporadic events of star formation, then their true number density is greater than the number density measured inside the selection box at any given redshift. The fraction of objects not observed due to this is also in principle calculable, with an analysis of this type and good, trustworthy modelling. 

In the case of the M10 models, this effect is much less important due to the smaller and more restricted amount of recent star formation. M10 models tend to be redder (especially in $r-i$), but mostly lie within the locus of observed galaxies, even when a small amounts of star formations is detected. 

Changes in colour tracks can not be attributed solely to the differences in the physical solutions. These were obtained by fitting the stacked spectra in the optical ($\lambda \gtrsim 3200 $\AA, depending on the models), and we compute the colours to a redshift range that surpasses the optical band, so UV extensions were used to accomplish this. Differences in the spectral libraries, and in the treatment of blue stragglers or blue horizontal branch stars affect the inferred colour evolution in this regime, even when if they have limited influence in the optical.

\subsection{Model comparison}\label{sec:goodness_of_fit}

\begin{figure}
  \begin{center}
   \includegraphics[width=3in]{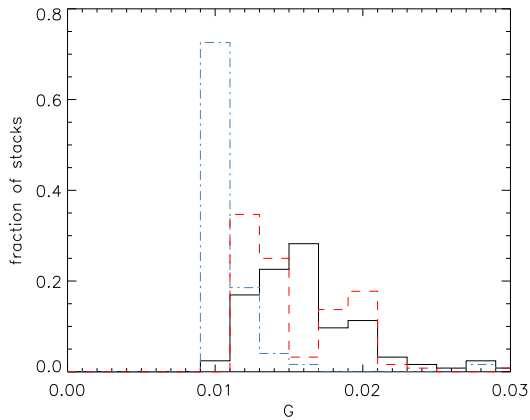}
    \caption{A simple measure of goodness of fit, taken as the average residual per pixel. In the black solid line we show the histogram for the results obtained with the BC03 models, in the red dashed line the results for the Mastro models, and in the blue dotted-dashed line the results for the FSPS models.}
    \label{fig:goodness_of_fit}
  \end{center}
\end{figure}

A natural question to ask is: which set of models best describes real galaxy spectra? Answering this question, however, is far from trivial. We can use the distribution of the average residual per pixel (as calculated in Section~\ref{sec:new_error} and shown for one stack in Fig.~\ref{fig:new_noise}), as the simplest possible measure of goodness of fit. We show the distribution of these values for our 124 stacks and for all three SPS models in Fig.~\ref{fig:goodness_of_fit}. There is a factor of roughly 1.5 between the mean residual of FSPS and BC03 or M10 models. This suggests FSPS models provide a better match overall for fitting the high SNR data. The physical solutions from this set of models are also the most in line with the LRG fiducial model, although the M10 solutions also broadly fit in with the mostly passive, metal-rich typical LRG picture. However the interpretation of this result is not straightforward and there are a number of caveats.

We are in a regime where neither set provides a statistically good fit to the data, and we do not know what the true answer should be. This means that when interpreting the best-fit solution and its residual when compared against the data, we need to bear in mind that the fitting procedure can compensate for deficiencies in the models by choosing a wrong solution. This behaviour can be explicitly seen in the ages of LRGs, where the best fit solution in data-space is known to be incorrect (see Section~\ref{sec:BB_prior} for more details). In that case, we have strong physical reason to impose a prior on the age of LRGs based on the age of the Universe obtained from independent methods. An approach based on strong physical priors, or on requiring different wavelength ranges to give consistent answers, can help to identify specific regions where each model is more likely to fail. These regions can in turn be interpreted in terms of stellar physics and library deficiencies, and be a useful tool for SPS models. Having a wider range of galaxy populations will also provide a broad test of the models. We leave this analysis for another publication.

\section{Summary and conclusions} \label{sec:conclusions}

We have computed, for the first time, colour-evolution models for the SDSS LRG sample, that depend on luminosity, colour and redshift, and are completely decoupled from the selection function of the survey. This is possible due to a full-spectral fitting technique (VESPA), which computes highly-resolved non-parametric star formation and metallicity histories, as well as the dust extinction. For 124 high signal-to-noise stacks of spectra for galaxies with different colour, luminosity and redshift, VESPA fits show how mass-weighted ages, metallicities, recent star formation and dust content vary. These solutions were then translated into colour-evolution tracks. For this analysis have considered three different SPS codes: BC03, M10 and FSPS, and studied the effect of this choice on the results.

Our results and conclusions can be summarised as follows:

\begin{enumerate}
\item When faced with high-quality data, the models are not able to produce formally good fits to the data. We use an updated error estimate, based on the fit residuals, as a way to deal with these limitations.
\item The broad picture that LRGs are dominated by old, metal-rich stellar populations is confirmed by our analysis using all SPS codes. However, whereas FSPS predicts virtually no recent star formation, M10 and BC03 require some amount of young stars to fit the spectra of the LRGs.
\item In the case of M10 and FSPS, we see some evidence for dependence of mass-weighted age on luminosity, with the faintest sample in both cases having a lower mass-weighted age. However, whereas in the M10 models this is due to star formation at young/intermediate ages, in the case of the FSPS models this is due to the dominant, old population being younger. Distinguishing between these two scenarios is crucial for studies that aim to use the age-redshift relationship of LRGs for cosmological constraints.
\item All three models reveal a similar picture for the dust extinction: extinction increases with decreasing luminosity, increasing redshift, and increasing $r-i$ colour.
\item Dust extinction is the only quantity we studied which has a significant and coherent gradient with $r-i$ colour.
\item The effect of recent star formation - even in small amounts - in the colour tracks can be quite dramatic, and temporarily remove LRGs from the SDSS LRG selection. It follows that, if LRGs really do go through such events, the number density of objects within the selection box is smaller than the true number density of LRGs. The fraction of LRGs going through this phase, as a function of redshift, can in principle be estimated using our models.
\item A test of which set of models best fits the data is non-trivial as no set gives a statistically good fit to the data. FSPS solutions produce, on average, lower residuals than the other two sets of models, but the significance of this cannot currently be established.
\item The differences in the colour tracks predicted by different SPS codes are the limiting factor in using these models to link populations of galaxies at different redshifts, and suggest that previous analyses that require this link (any based on the evolution of a quantity with redshift, such as number or luminosity densities, clustering, etc) are likely to at best have underestimated their errors, and at worse suffered from a systematic error due to our lack of understanding in stellar evolution.
\end{enumerate}

Lastly, we note that the results in this paper apply only to the LRG population, which is a subset of the ETG population. The relationship is more pronounced in terms of mass than colour - the LRG population occupies the most massive end of the ETG population, which is in itself predominantly red. 

The colour evolution tracks obtained with any of the SPS models can be found in \url{http://www.icg.port.ac.uk/~tojeiror/lrg_evolution/.}

\section{Acknowledgments}
We thank the anonymous referee for a thoughtful report that made this manuscript noticeably clearer. We would like to thank Claudia Maraston and Charlie Conroy for helpful discussions and for providing UV extensions of their models and comments on an earlier draft. We also thank David Wake for helpful and encouraging discussions.
 
RT thanks the Leverhulme trust for financial support. WJP is
grateful for support from the UK Science and Technology Facilities
Council, the Leverhulme trust and the European Research Council. 

    Funding for the SDSS and SDSS-II has been provided by the Alfred
    P. Sloan Foundation, the Participating Institutions, the National
    Science Foundation, the U.S. Department of Energy, the National
    Aeronautics and Space Administration, the Japanese Monbukagakusho,
    the Max Planck Society, and the Higher Education Funding Council
    for England. The SDSS Web Site is http://www.sdss.org/. The SDSS is managed by the Astrophysical Research Consortium for the Participating Institutions. The Participating Institutions are the American Museum of Natural History, Astrophysical Institute Potsdam, University of Basel, University of Cambridge, Case Western Reserve University, University of Chicago, Drexel University, Fermilab, the Institute for Advanced Study, the Japan Participation Group, Johns Hopkins University, the Joint Institute for Nuclear Astrophysics, the Kavli Institute for Particle Astrophysics and Cosmology, the Korean Scientist Group, the Chinese Academy of Sciences (LAMOST), Los Alamos National Laboratory, the Max-Planck-Institute for Astronomy (MPIA), the Max-Planck-Institute for Astrophysics (MPA), New Mexico State University, Ohio State University, University of Pittsburgh, University of Portsmouth, Princeton University, the United States Naval Observatory, and the University of Washington.

\bibliographystyle{mn2e}
\bibliography{/Users/ritat//WORK/LRG/PAPER/my_bibliography}

\appendix
\section[]{Additional plots with different SPS models}

In this Appendix we show the equivalent of the plots shown in the main text, for all other SPS models considered. All plots are like for like, and we refer the reader to the main text for an explanation.

\subsection{Fits and residuals} \label{sec:appendix_fits}

\begin{figure}
  \begin{center}
   \includegraphics[width=3.5in]{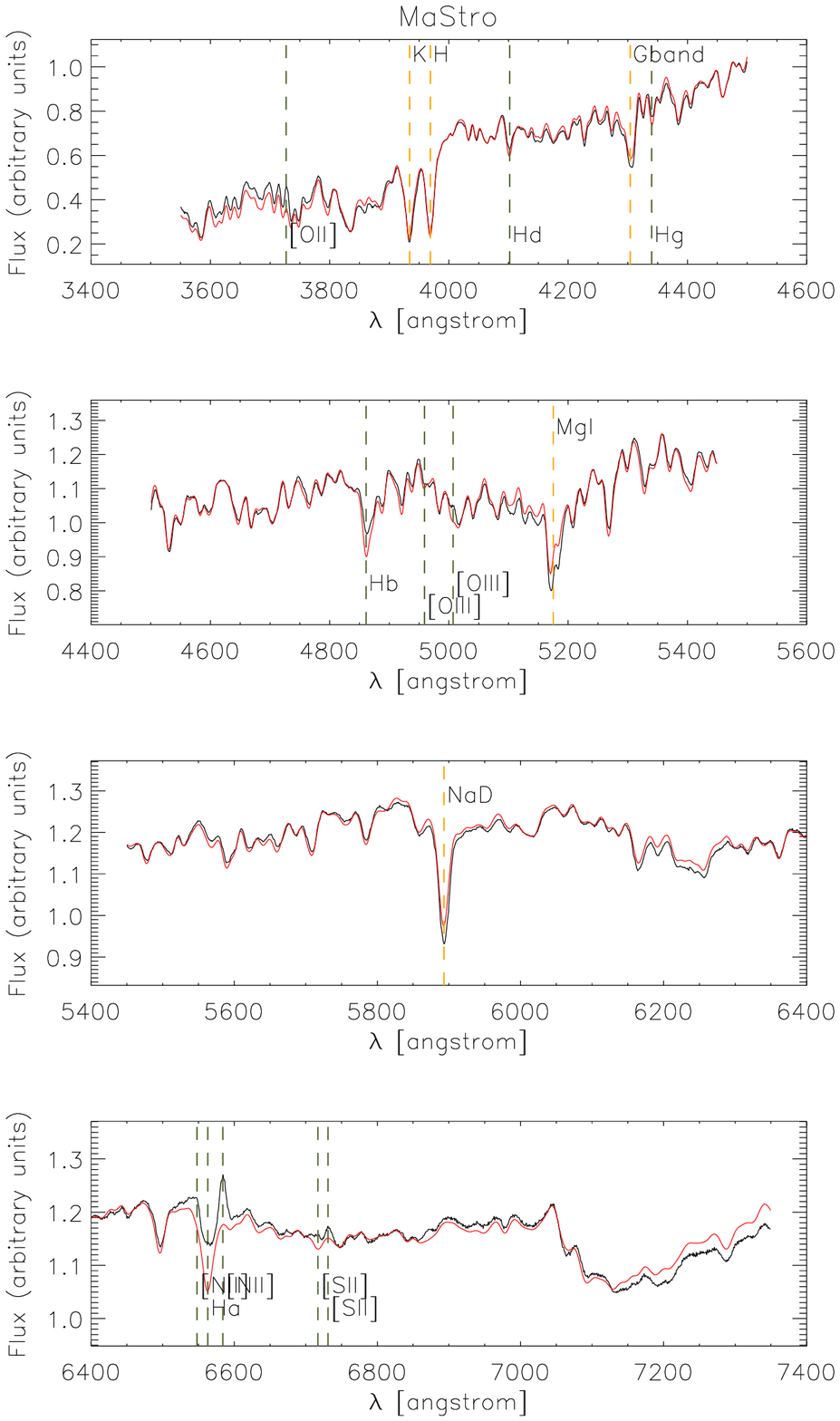}
    \caption{A typical fit, using the M10 models. The black line is the data, and the red line the best-fitting model. The vertical yellow and green dashed lines guide the eye by showing some common emission and absorption features. Although plotted, some regions of the spectrum were excluded from the fit - see Fig.~\ref{fig:residuals_M10}.}
    \label{fig:fit_M10}
  \end{center}
\end{figure}

\begin{figure}
  \begin{center}
   \includegraphics[width=3.5in]{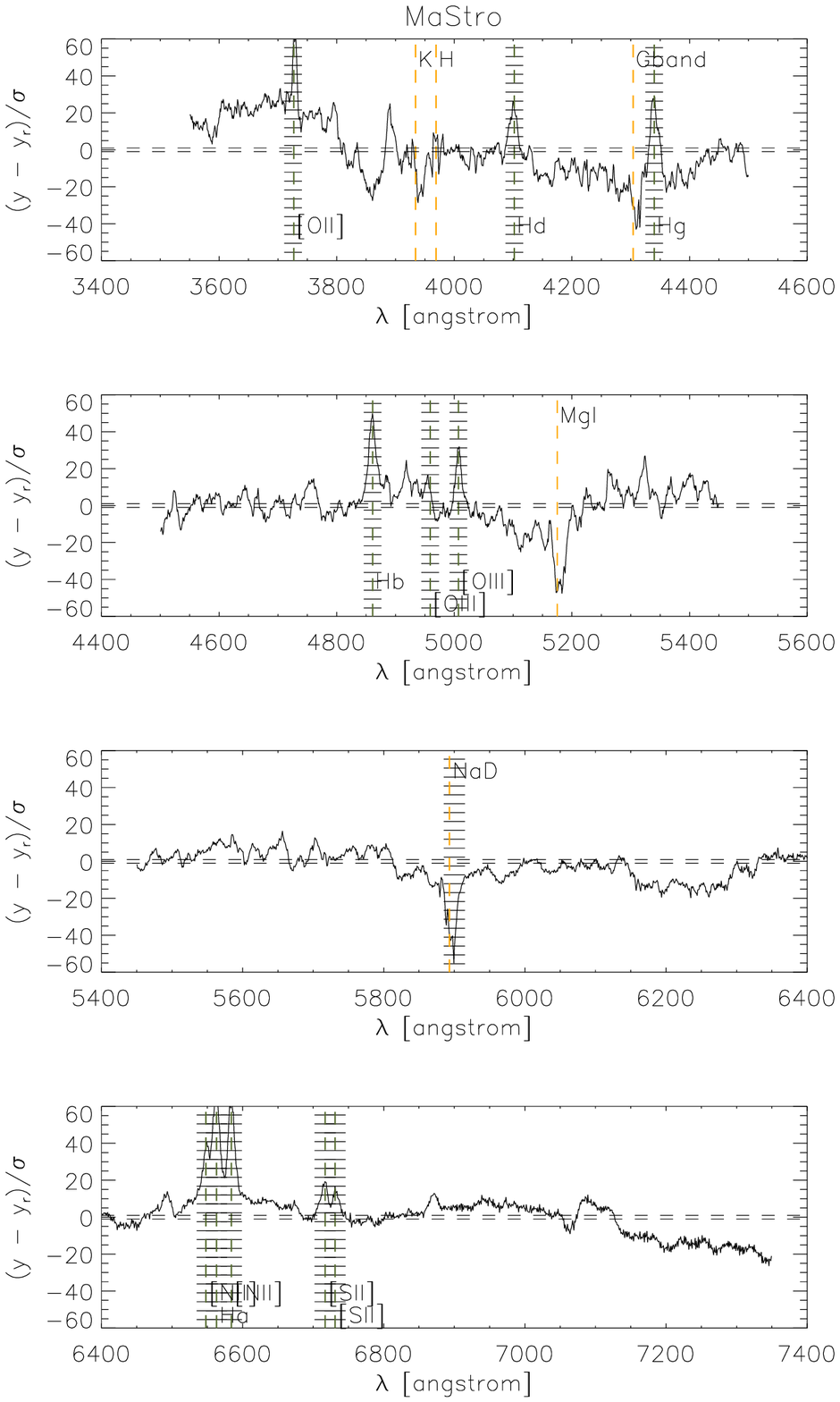}
    \caption{The residuals in units of the stack noise, for the fit showed in Fig.~\ref{fig:fit_M10}. The vertical dashed lines guide the eye by showing some common emission and absorption features. The regions shaded by horizontal lines were excluded from the fit.}
    \label{fig:residuals_M10}
  \end{center}
\end{figure}

\begin{figure}
  \begin{center}
   \includegraphics[width=3.5in]{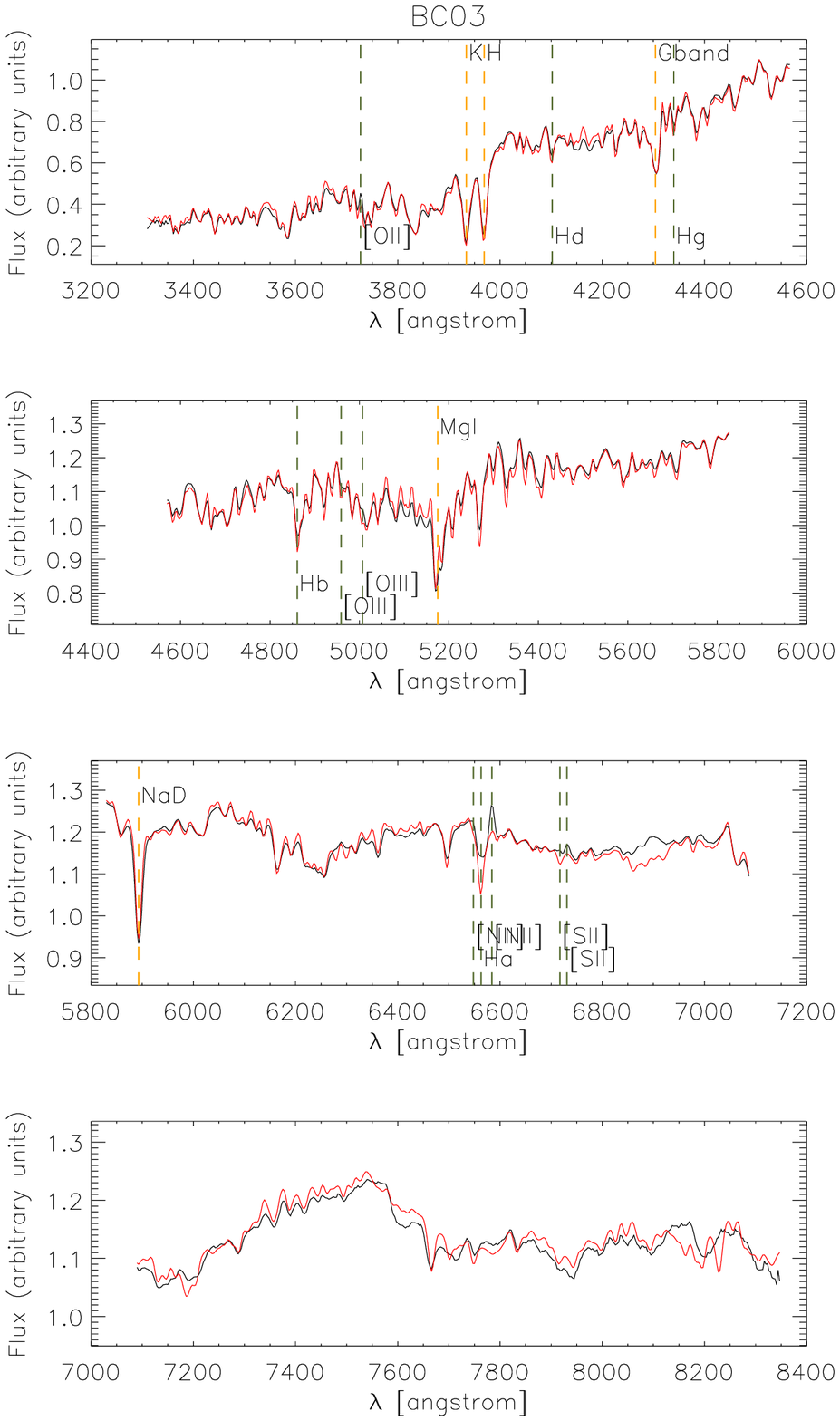}
    \caption{A typical fit, using the BC03 models. The black line is the data, and the red line the best-fitting model. The vertical yellow and green dashed lines guide the eye by showing some common emission and absorption features. Although plotted, some regions of the spectrum were excluded from the fit - see Fig.~\ref{fig:residuals_BC03}.}
    \label{fig:fit_BC03}
  \end{center}
\end{figure}

In this section we show the equivalent of Fig.~\ref{fig:test_fit} and Fig.~\ref{fig:test_Residuals} for the same stacked spectra, analysed with the M10 models (in Fig.~\ref{fig:fit_M10} and Fig.~\ref{fig:residuals_M10}) and with the BC03 models (in Fig.~\ref{fig:fit_BC03} and Fig.~\ref{fig:fit_BC03}).

\begin{figure}
  \begin{center}
   \includegraphics[width=3.5in]{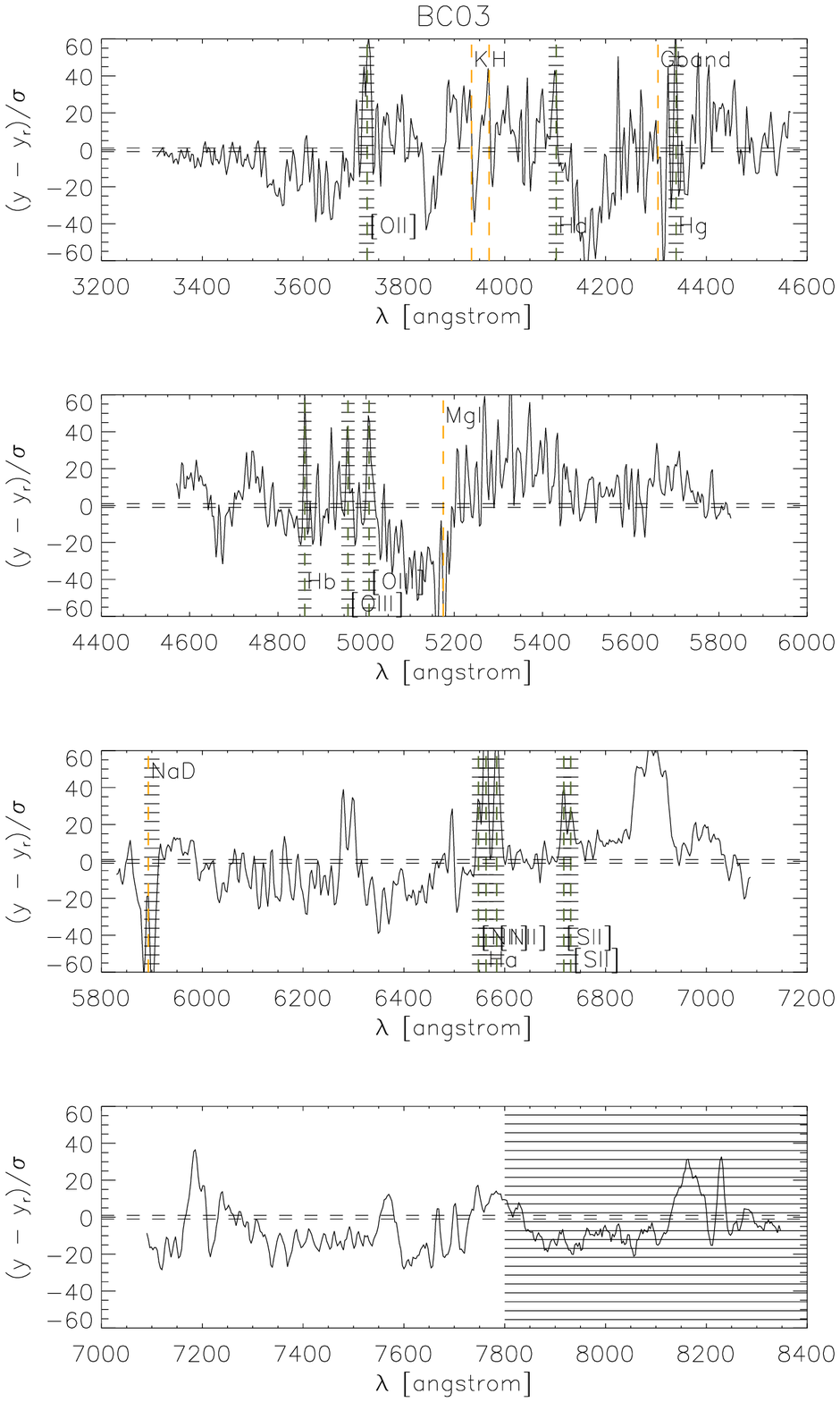}
    \caption{The residuals in units of the stack noise, for the fit showed in Fig.~\ref{fig:fit_BC03}. The vertical dashed lines guide the eye by showing some common emission and absorption features. The regions shaded by horizontal lines were excluded from the fit.}
    \label{fig:residuals_BC03}
  \end{center}
\end{figure}

\subsection{2D histograms}\label{sec:appendix_2dhists}

\begin{figure*}
  \begin{center}
   \includegraphics[width=3.5in, angle=90]{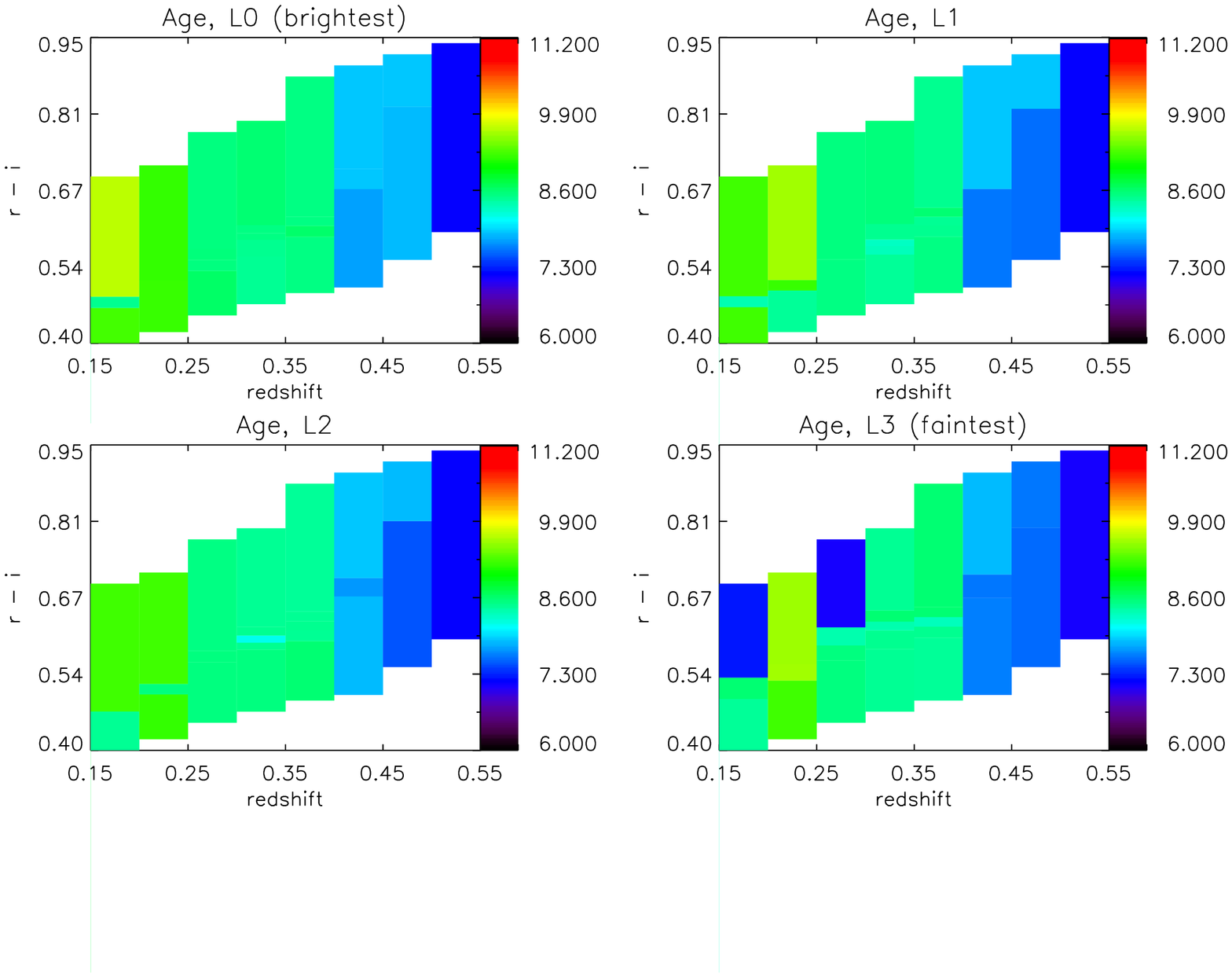}
    \caption{Mass-weighted age in Gyr (see Eq.~\ref{eq:mass_weighted_age}) for galaxies of different colour, redshift and rest-frame $r-$band luminosity, analysed with the FSPS models. We only show data for regions of parameter space with a sufficient number density of galaxies. In practice, there are a few galaxies (less than 20 per redshift bin) outside of the coloured areas, but these have an insignificant impact on the recovered solution. In Fig.~\ref{fig:alltrends_allmodels} (3rd row) we average over colour to show the trend with redshift for each luminosity slice.}
    \label{fig:age_2dhist_FSPS}
  \end{center}
\end{figure*}

\begin{figure*}
  \begin{center}
   \includegraphics[width=3.5in, angle=90]{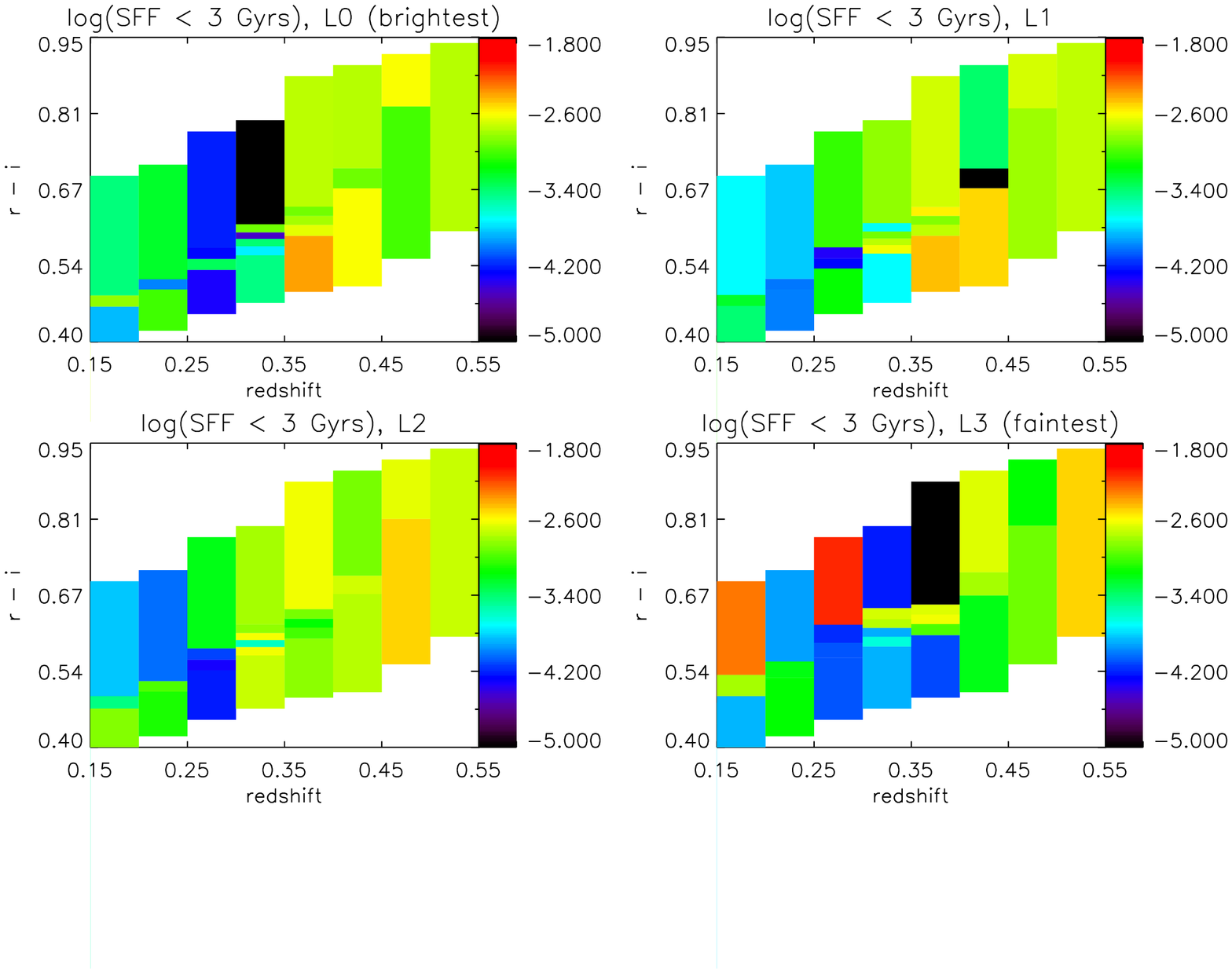}
    \caption{The logarithm of the fraction of star formation (by mass), recovered in bins up to 3.8 Gyr for galaxies of different colour, redshift and rest-frame $r-$band luminosity, analysed with the FSPS models. We only show data for regions of parameter space with a sufficient number density of galaxies. In practice, there are a few galaxies (less than 20 per redshift bin) outside of the coloured areas, but these have an insignificant impact on the recovered solution. In Fig.~\ref{fig:alltrends_allmodels} (5th row) we average over colour to show the trend with redshift for each luminosity slice.}
    \label{fig:RSF_2dhist_FSPS}
  \end{center}
\end{figure*}

\begin{figure*}
  \begin{center}
   \includegraphics[width=3.5in, angle=90]{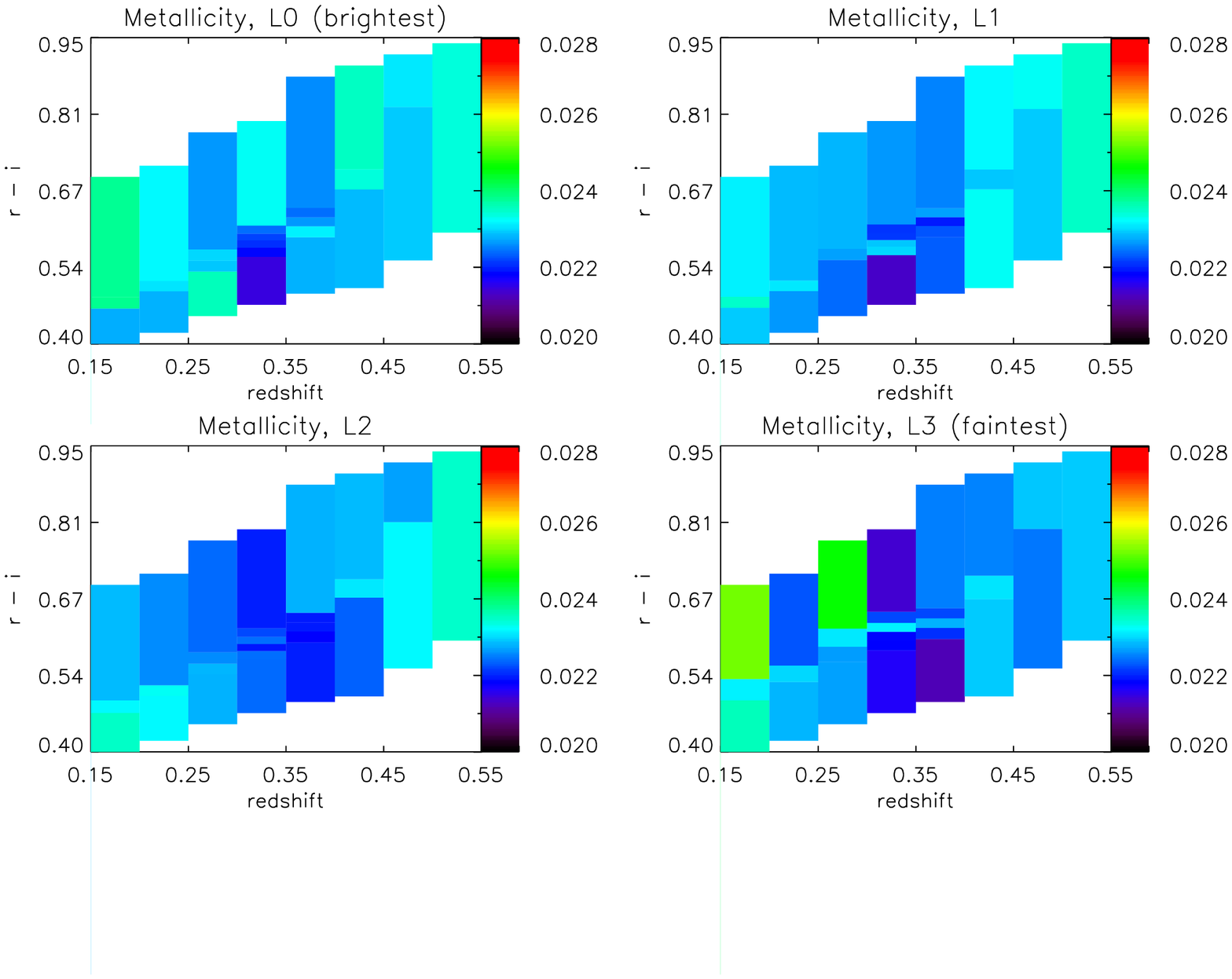}
    \caption{Mass-weighted metallicity (see Eq.~\ref{eq:mass_weighted_metallicity}) for galaxies of different colour, redshift and rest-frame $r-$band luminosity, analysed with the FSPS models. We only show data for regions of parameter space with a sufficient number density of galaxies. In practice, there are a few galaxies (less than 20 per redshift bin) outside of the coloured areas, but these have an insignificant impact on the recovered solution. Fig.~\ref{fig:alltrends_allmodels} averages over colour to show the trend with redshift for each luminosity slice.}
    \label{fig:metallicity_2dhist_FSPS}
  \end{center}
\end{figure*}

\begin{figure*}
  \begin{center}
   \includegraphics[width=3.5in, angle=90]{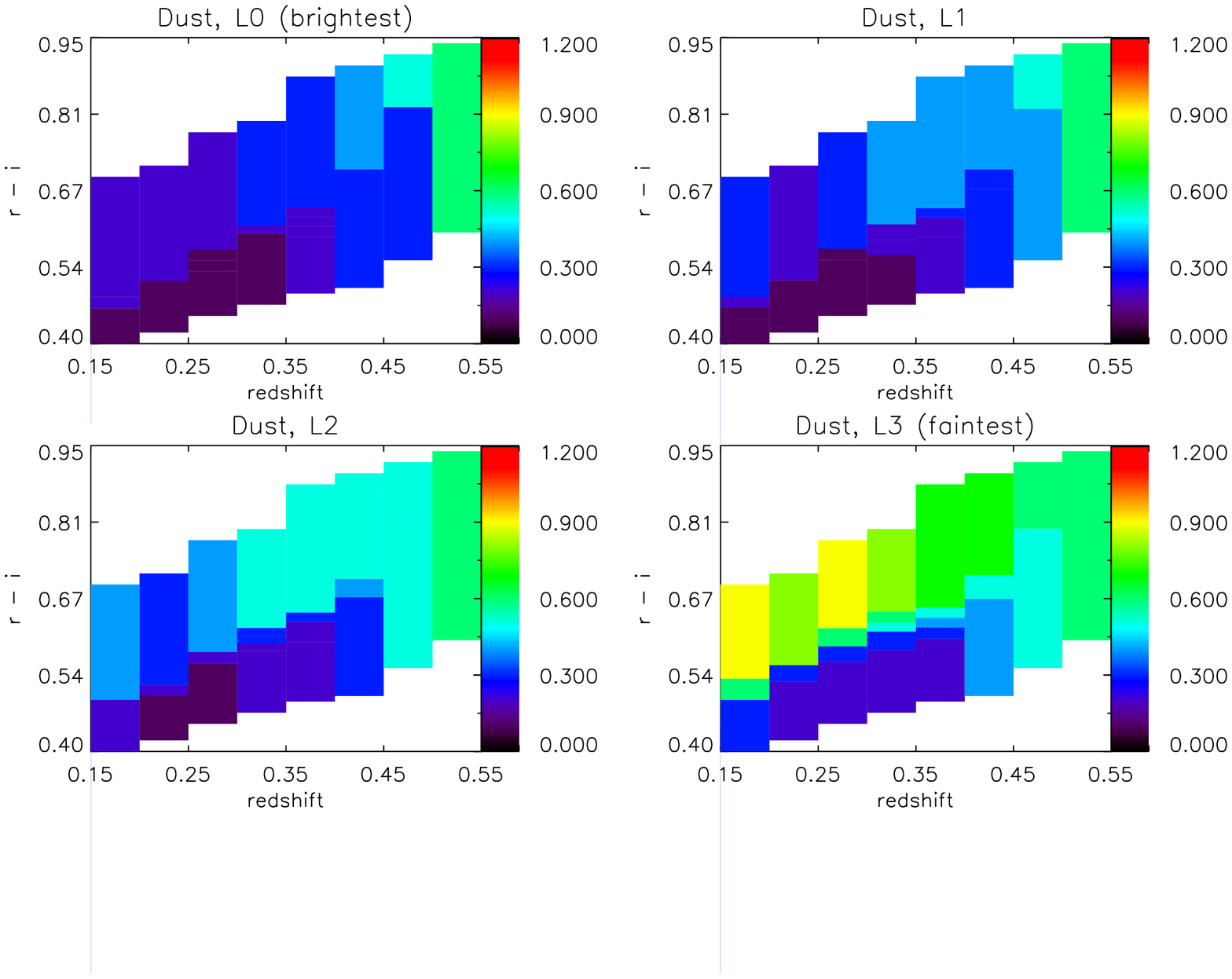}
    \caption{Inter-stellar dust absorption for galaxies of different colour, redshift and rest-frame $r-$band luminosity, analysed with the FSPS models. We only show data for regions of parameter space with a sufficient number density of galaxies. In practice, there are a few galaxies (less than 20 per redshift bin) outside of the coloured areas, but these have an insignificant impact on the recovered solution. Fig.~\ref{fig:alltrends_allmodels} averages over colour to show the trend with redshift for each luminosity slice.}
    \label{fig:dust_2dhist_FSPS}
  \end{center}
\end{figure*}

\begin{figure*}
  \begin{center}
   \includegraphics[width=3.5in, angle=90]{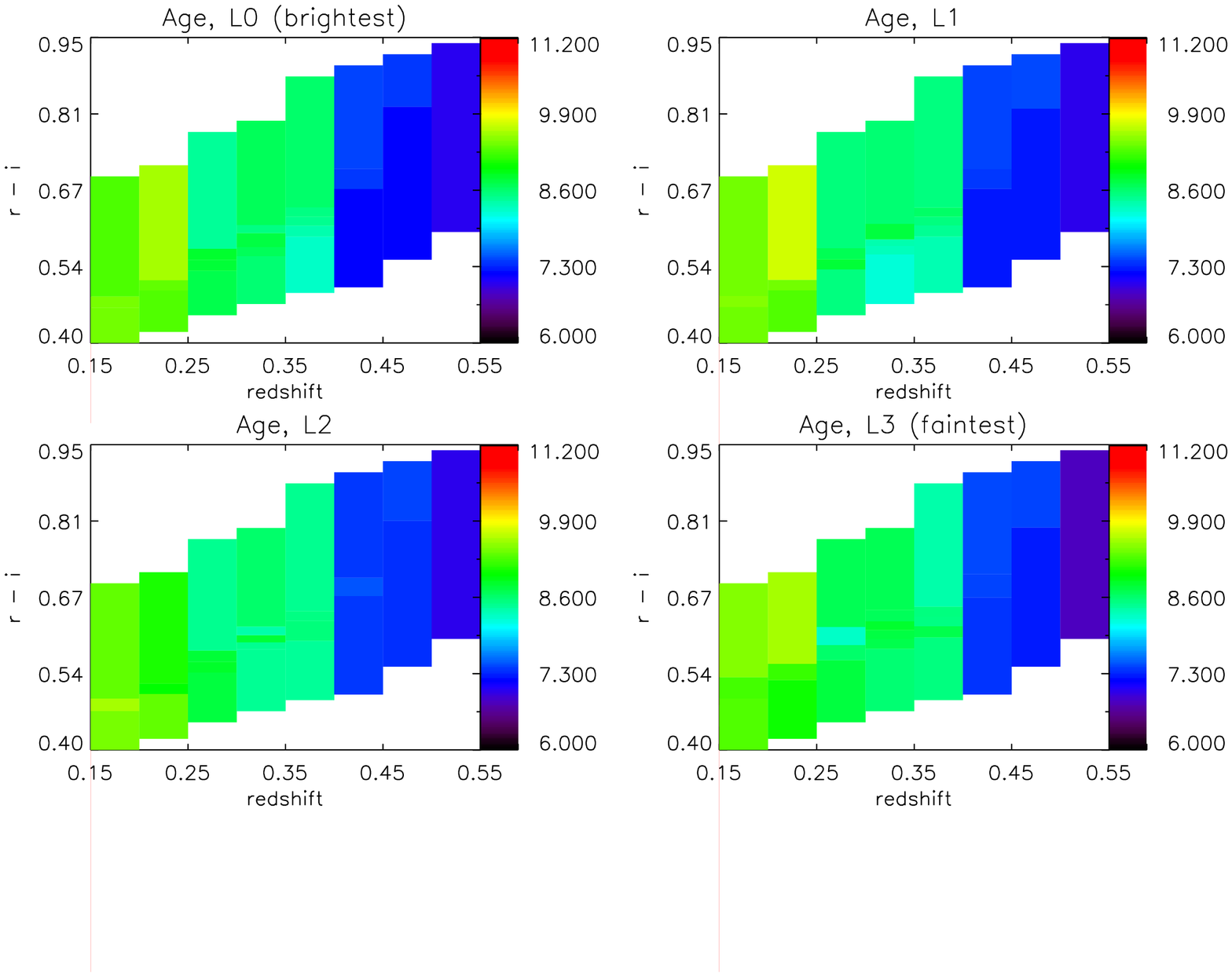}
    \caption{Mass-weighted age in Gyr (see Eq.~\ref{eq:mass_weighted_age}) for galaxies of different colour, redshift and rest-frame $r-$band luminosity, analysed with the BC03 models. We only show data for regions of parameter space with a sufficient number density of galaxies. In practice, there are a few galaxies (less than 20 per redshift bin) outside of the coloured areas, but these have an insignificant impact on the recovered solution. In Fig.~\ref{fig:alltrends_allmodels} (3rd row) we average over colour to show the trend with redshift for each luminosity slice.}
    \label{fig:age_2dhist_BC03}
  \end{center}
\end{figure*}

\begin{figure*}
  \begin{center}
   \includegraphics[width=3.5in, angle=90]{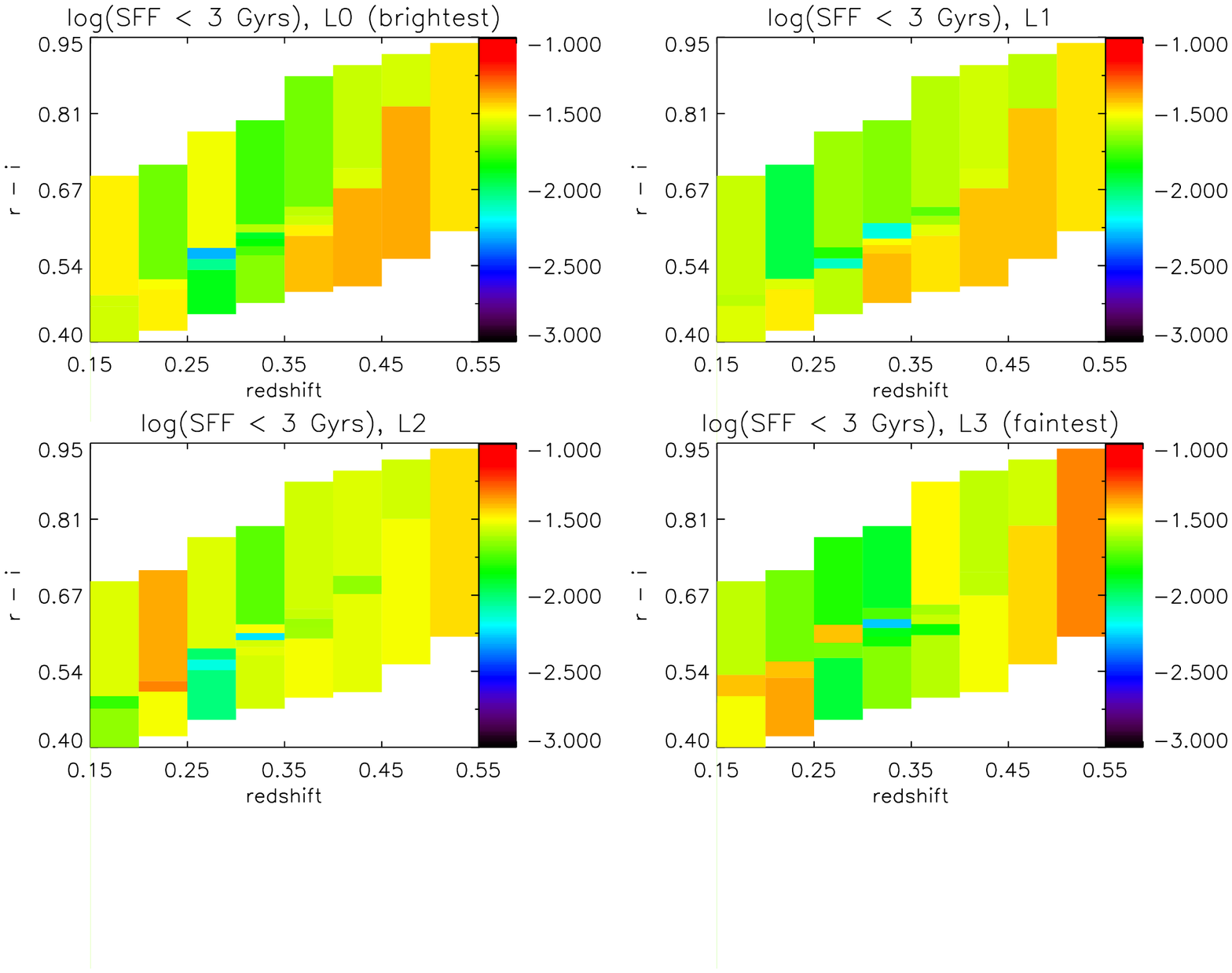}
    \caption{The logarithm of the fraction of star formation (by mass), recovered in bins up to 3.8 Gyr for galaxies of different colour, redshift and rest-frame $r-$band luminosity, analysed with the BC03 models. We only show data for regions of parameter space with a sufficient number density of galaxies. In practice, there are a few galaxies (less than 20 per redshift bin) outside of the coloured areas, but these have an insignificant impact on the recovered solution. In Fig.~\ref{fig:alltrends_allmodels} (5th row) we average over colour to show the trend with redshift for each luminosity slice.}
    \label{fig:RSF_2dhist_BC03}
  \end{center}
\end{figure*}

\begin{figure*}
  \begin{center}
   \includegraphics[width=3.5in, angle=90]{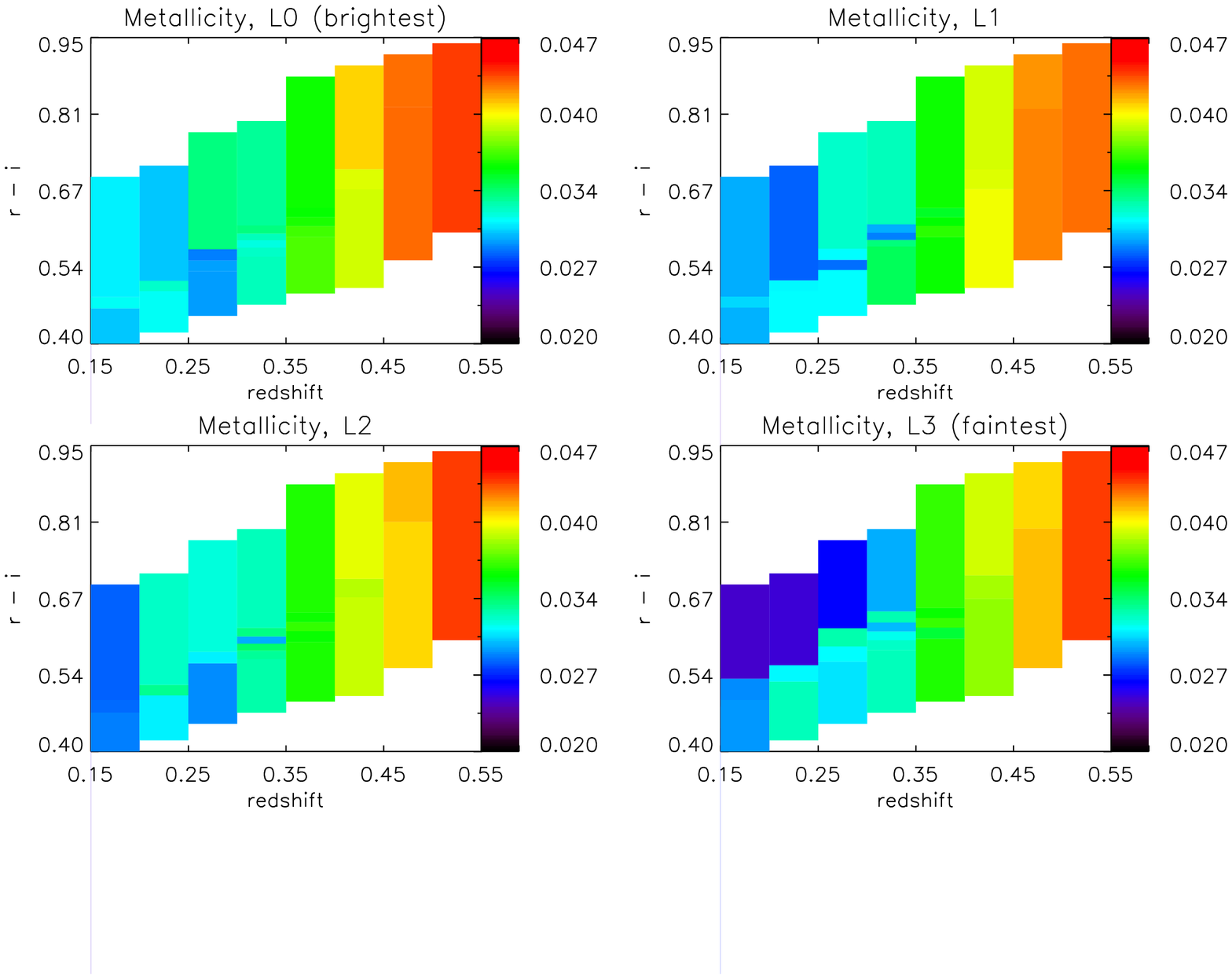}
    \caption{Mass-weighted metallicity (see Eq.~\ref{eq:mass_weighted_metallicity}) for galaxies of different colour, redshift and rest-frame $r-$band luminosity, analysed with the BC03. We only show data for regions of parameter space with a sufficient number density of galaxies. In practice, there are a few galaxies (less than 20 per redshift bin) outside of the coloured areas, but these have an insignificant impact on the recovered solution. Fig.~\ref{fig:alltrends_allmodels} averages over colour to show the trend with redshift for each luminosity slice.}
    \label{fig:metallicity_2dhist_BC03}
  \end{center}
\end{figure*}

\begin{figure*}
  \begin{center}
   \includegraphics[width=3.5in, angle=90]{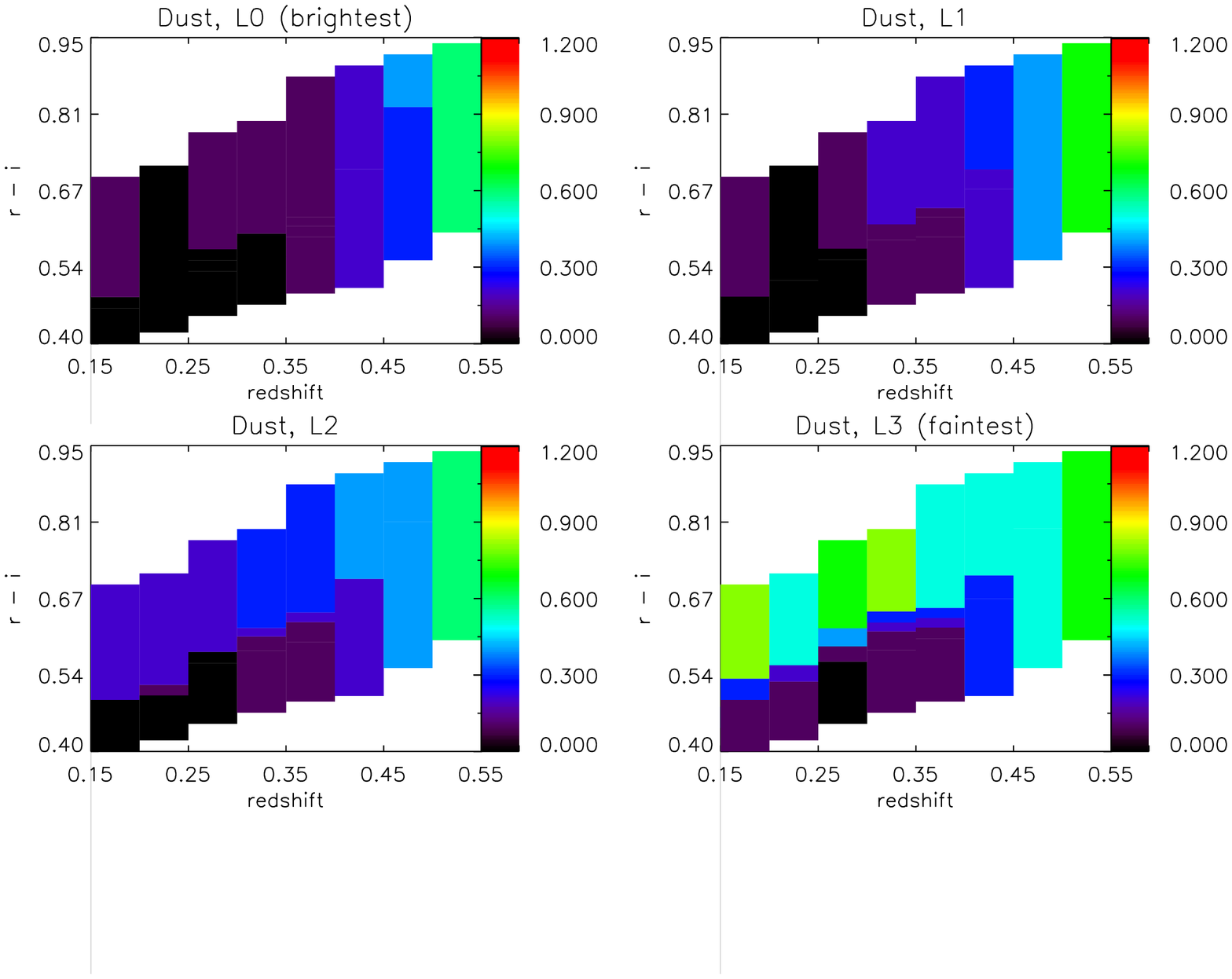}
    \caption{Inter-stellar dust absorption for galaxies of different colour, redshift and rest-frame $r-$band luminosity, analysed with the BC03 models. We only show data for regions of parameter space with a sufficient number density of galaxies. In practice, there are a few galaxies (less than 20 per redshift bin) outside of the coloured areas, but these have an insignificant impact on the recovered solution. Fig.~\ref{fig:alltrends_allmodels} averages over colour to show the trend with redshift for each luminosity slice.}
    \label{fig:dust_2dhist_BC03}
  \end{center}
\end{figure*}

In this section we present the 2D histograms of Section~\ref{sec:measured_quantities}, which show the dependence of age, metallicity, dust extinction and amount of recent star formation ($< 3$ Gyr) as a function of colour and redshift. Figs.~\ref{fig:age_2dhist_FSPS} to~\ref{fig:dust_2dhist_FSPS} show the results obtained with using the FSPS code, and Figs.~\ref{fig:age_2dhist_BC03} to~\ref{fig:dust_2dhist_BC03} the results obtained using the BC03 code.

\subsection{Colour tracks} \label{sec:appendix_colour_tracks}

\begin{figure}
  \begin{center}
   \includegraphics[width=3.5in]{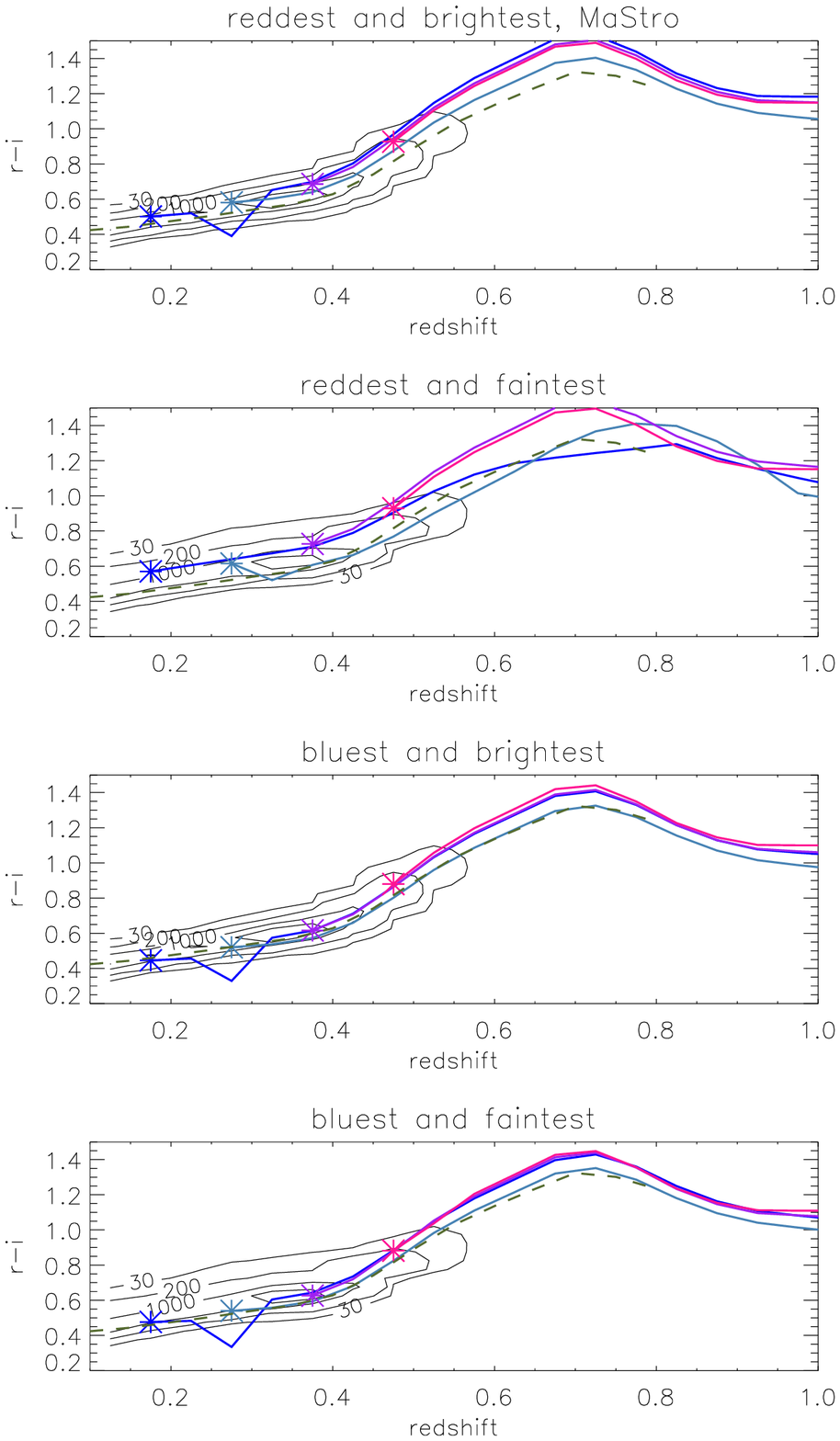}
    \caption{Predicted evolution of $r-i$ colour with redshift, calculated according to Eq.~(\ref{eq:F_te}), and the solutions presented in Section~\ref{sec:measured_quantities}, and using the M10. The different panels show examples at different combinations of colour and luminosity. The black contour lines show the number density of LRGs within the respective luminosity slice, and the coloured lines show the tracks from cells at four different redshifts: 0.175, 0.275, 0.375 and 0.475. The star shows the position of the cell the track relates to. The green dashed line shows the LRG model from \citet{MarastonEtAl09}, for comparison.}
    \label{fig:rmi_tracks_M10}
  \end{center}
\end{figure}

\begin{figure}
  \begin{center}
   \includegraphics[width=3.5in]{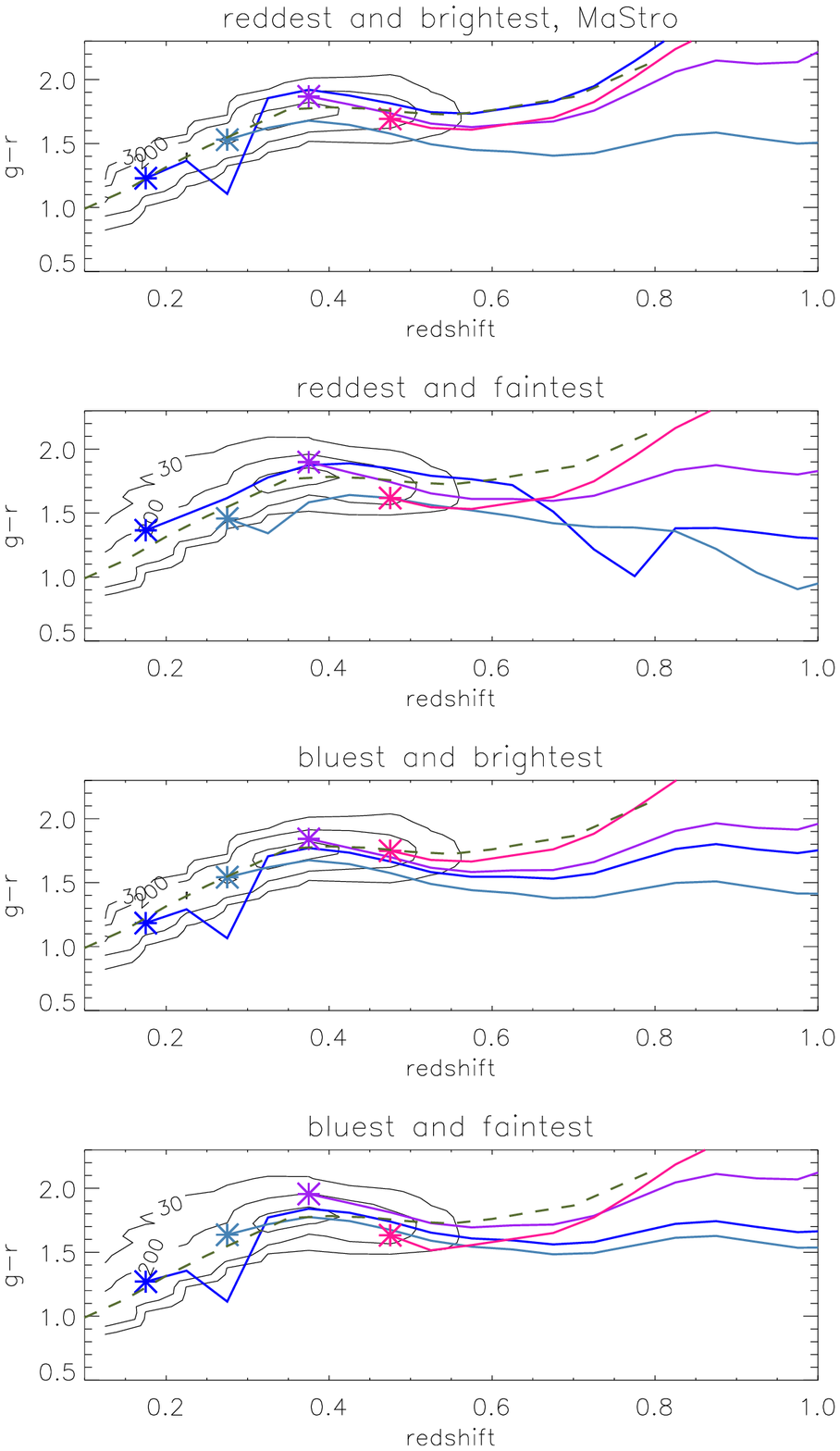}
    \caption{Predicted evolution of $g-r$ colour with redshift, calculated according to Eq.~(\ref{eq:F_te}), and the solutions presented in Section~\ref{sec:measured_quantities}, and using the M10. The different panels show examples at different combinations of colour and luminosity. The black contour lines show the number density of LRGs within the respective luminosity slice, and the coloured lines show the tracks from cells at four different redshifts: 0.175, 0.275, 0.375 and 0.475. The star shows the position of the cell the track relates to. The green dashed line shows the LRG model from \citet{MarastonEtAl09}, for comparison.}
    \label{fig:gmr_tracks_M10}
  \end{center}
\end{figure}

\begin{figure}
  \begin{center}
   \includegraphics[width=3.5in]{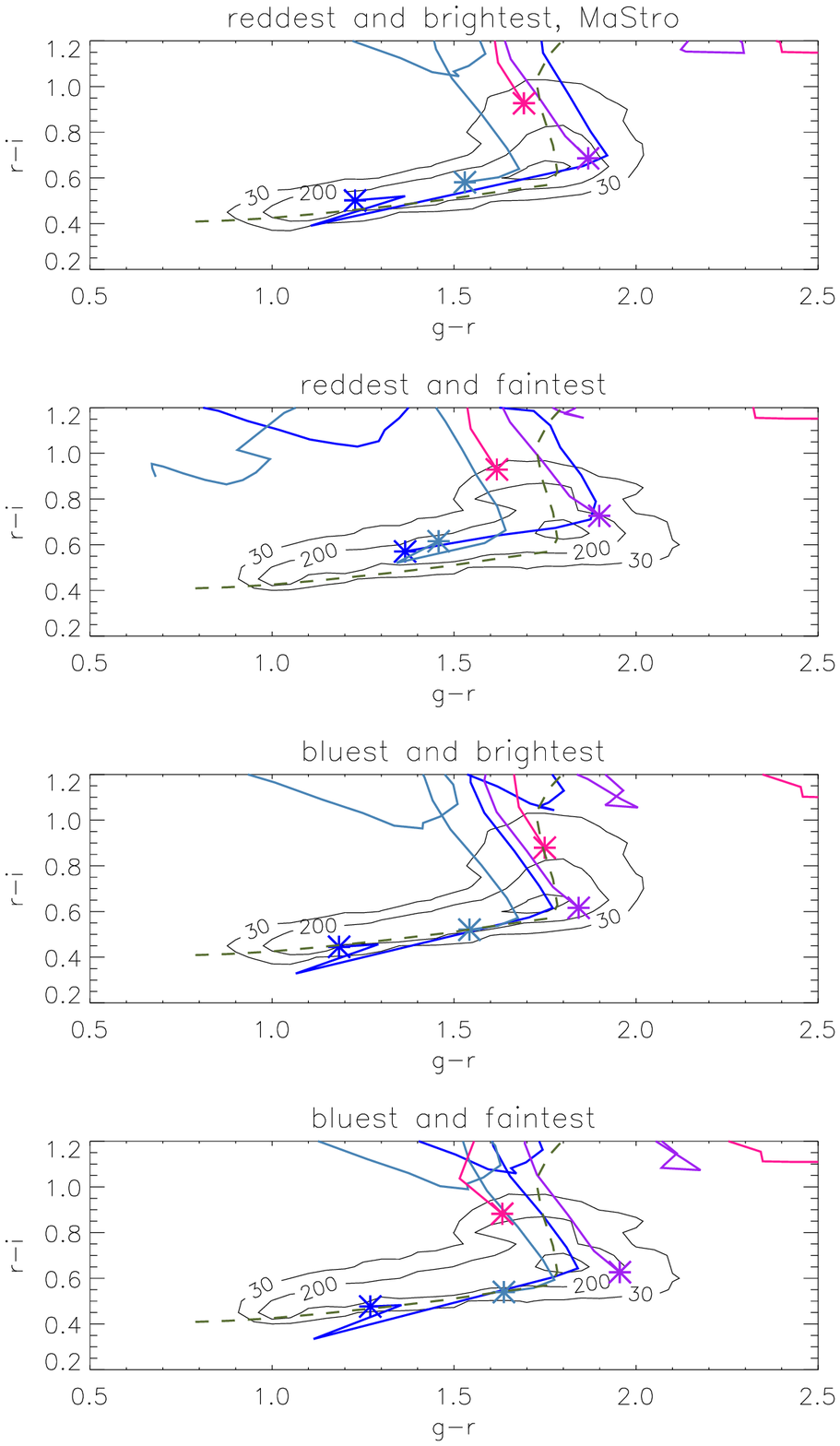}
    \caption{Predicted evolution of $r-i$ colour with $g-r$, calculated according to Eq.~(\ref{eq:F_te}), and the solutions presented in Section~\ref{sec:measured_quantities}, and using the M10 models. The different panels show examples at different combinations of colour and luminosity. The black contour lines show the number density of LRGs within the respective luminosity slice, and the coloured lines show the tracks from cells at four different redshifts: 0.175, 0.275, 0.375 and 0.475. The star shows the position of the cell the track relates to. The green dashed line shows the LRG model from \citet{MarastonEtAl09}, for comparison.}
    \label{fig:rmi_gmr_tracks_M10}
  \end{center}
\end{figure}

\begin{figure}
  \begin{center}
   \includegraphics[width=3.5in]{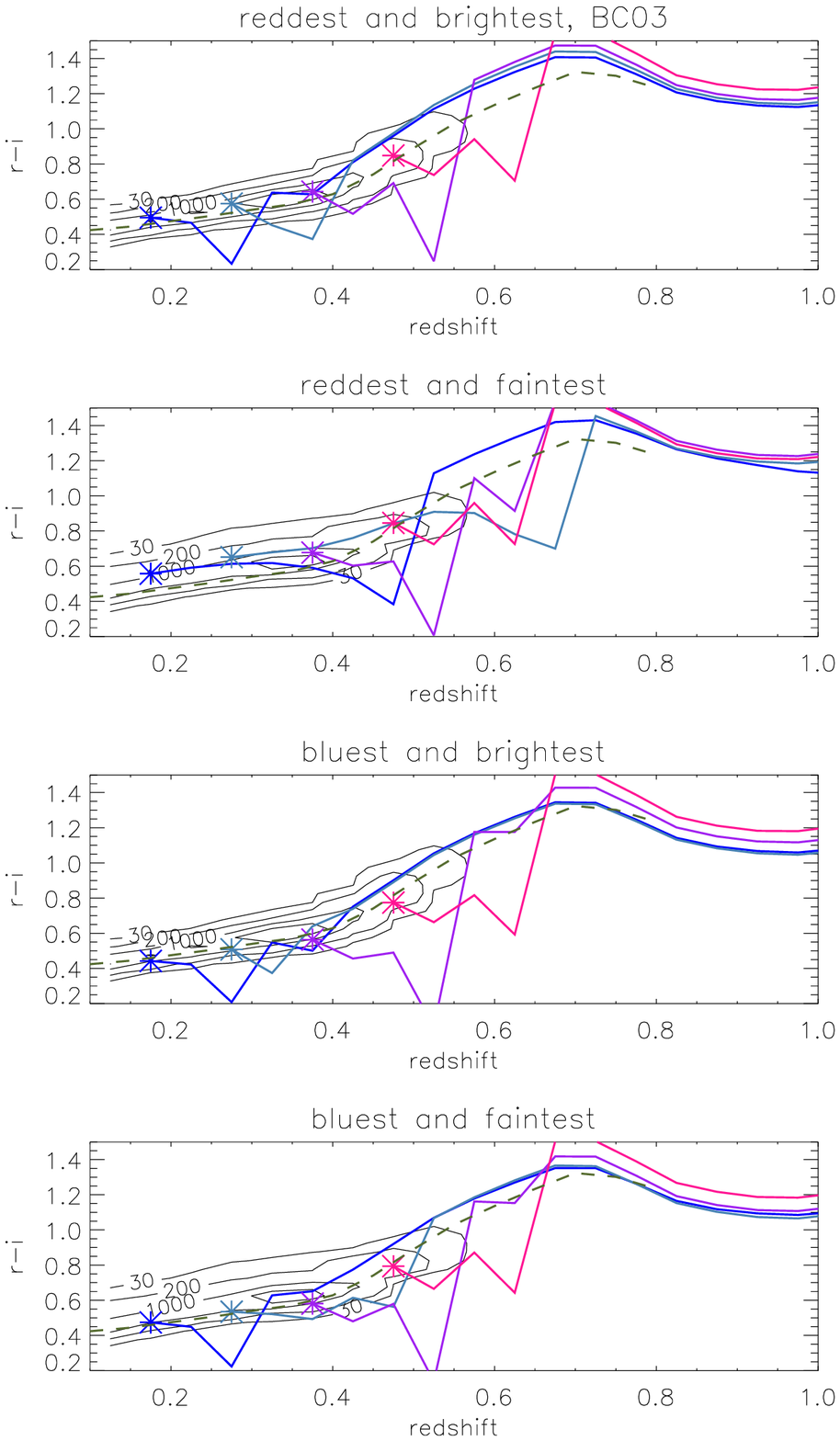}
    \caption{Predicted evolution of $r-i$ colour with redshift, calculated according to Eq.~(\ref{eq:F_te}), and the solutions presented in Section~\ref{sec:measured_quantities}, and using the BC03 models. The different panels show examples at different combinations of colour and luminosity. The black contour lines show the number density of LRGs within the respective luminosity slice, and the coloured lines show the tracks from cells at four different redshifts: 0.175, 0.275, 0.375 and 0.475. The star shows the position of the cell the track relates to. The green dashed line shows the LRG model from \citet{MarastonEtAl09}, for comparison.}
    \label{fig:rmi_tracks_BC03}
  \end{center}
\end{figure}

\begin{figure}
  \begin{center}
   \includegraphics[width=3.5in]{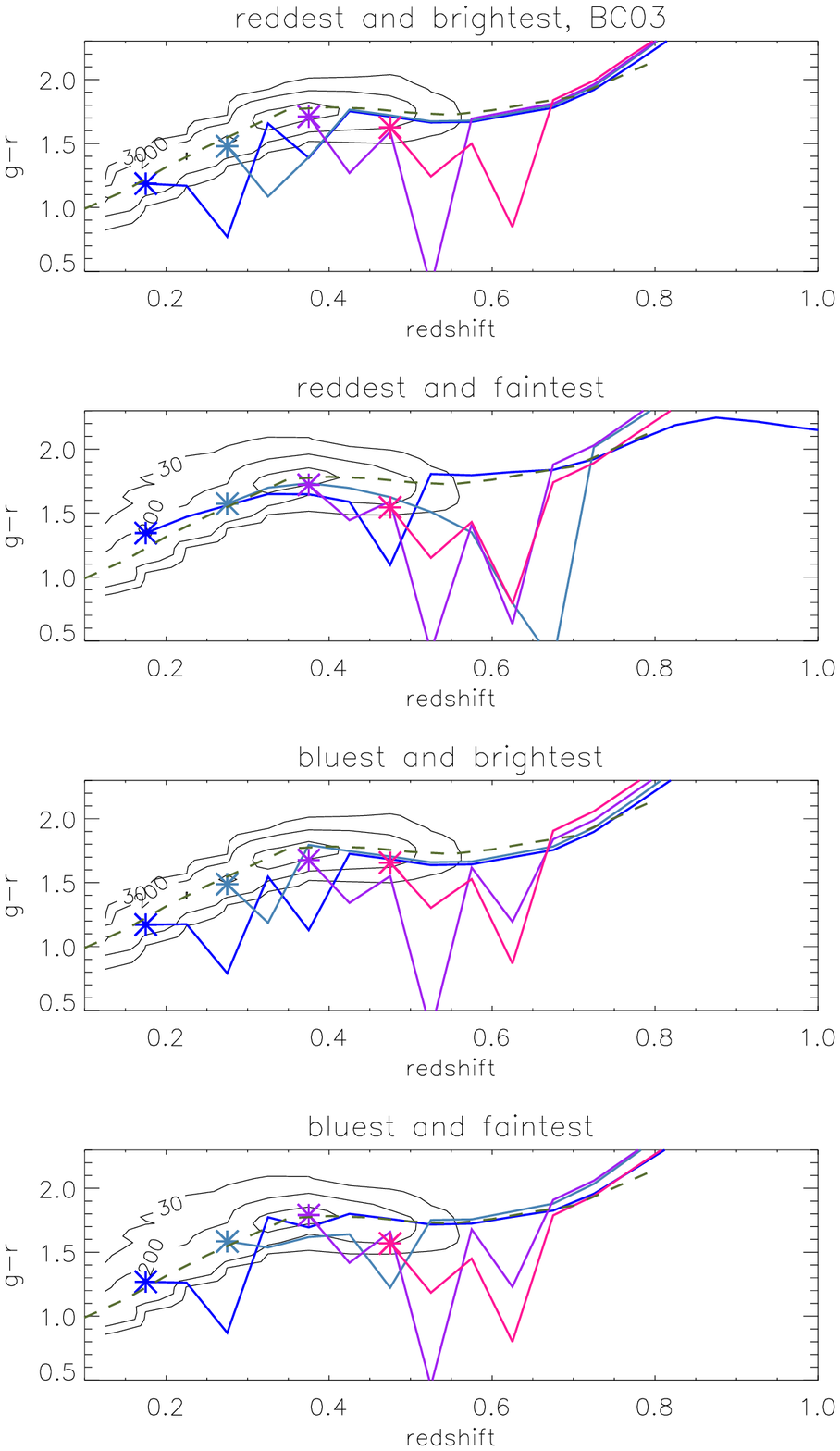}
    \caption{Predicted evolution of $g-r$ colour with redshift, calculated according to Eq.~(\ref{eq:F_te}), and the solutions presented in Section~\ref{sec:measured_quantities}, and using the BC03 models. The different panels show examples at different combinations of colour and luminosity. The black contour lines show the number density of LRGs within the respective luminosity slice, and the coloured lines show the tracks from cells at four different redshifts: 0.175, 0.275, 0.375 and 0.475. The star shows the position of the cell the track relates to. The green dashed line shows the LRG model from \citet{MarastonEtAl09}, for comparison.}
    \label{fig:gmr_tracks_BC03}
  \end{center}
\end{figure}

\begin{figure}
  \begin{center}
   \includegraphics[width=3.5in]{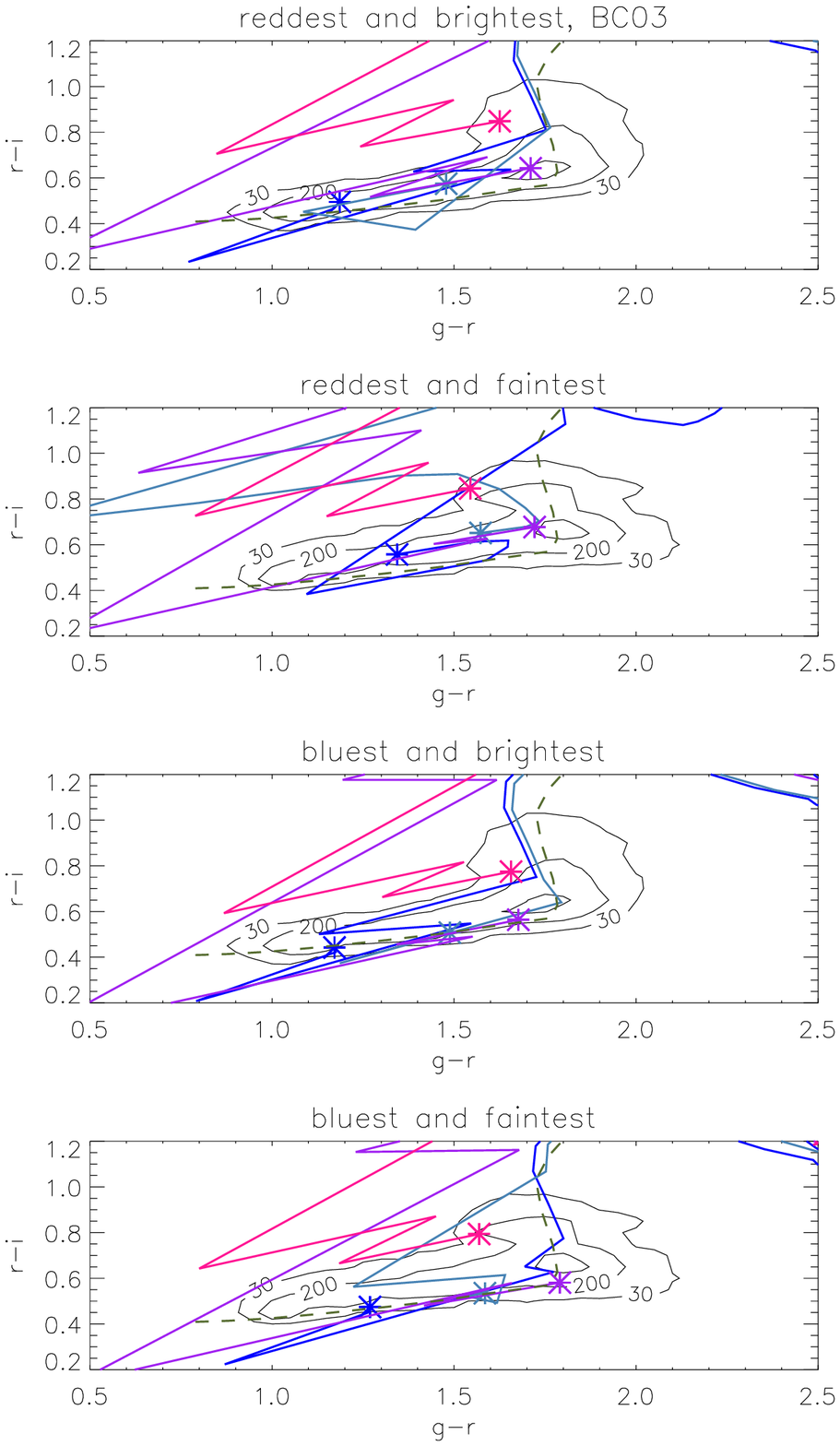}
    \caption{Predicted evolution of $r-i$ colour with $g-r$, calculated according to Eq.~(\ref{eq:F_te}), and the solutions presented in Section~\ref{sec:measured_quantities}, and using the BC03 models. The different panels show examples at different combinations of colour and luminosity. The black contour lines show the number density of LRGs within the respective luminosity slice, and the coloured lines show the tracks from cells at four different redshifts: 0.175, 0.275, 0.375 and 0.475. The star shows the position of the cell the track relates to. The green dashed line shows the LRG model from \citet{MarastonEtAl09}, for comparison.}
    \label{fig:rmi_gmr_tracks_BC03}
  \end{center}
\end{figure}

Here we present the colour tracks, as in shown Section~\ref{sec:colour_tracks}, computed using the M10 models (Figs.~\ref{fig:rmi_tracks_M10} to~\ref{fig:rmi_gmr_tracks_M10}) and the BC03 models (Figs.~\ref{fig:rmi_tracks_BC03} to~\ref{fig:rmi_gmr_tracks_BC03}).

\end{document}